\documentclass[12pt,preprint]{aastex}

\usepackage{emulateapj5}
\usepackage{natbib,apjfonts,graphicx}

\newcommand{\chandra}{{\sl Chandra}~}

\newcommand{\mdot}{ \dot{M} }

\newcommand{\yr}{{\,\rm yr\,}}
\newcommand{\kms}{{\mbox{$\,\rm km~s^{-1}$}}}
\newcommand{\msun}{\,M_{\sun}\,}

\def\mathfont#1{\ifmmode{#1}\else{$#1$}\fi} 
\def\lae{\mathrel{<\kern-1.0em\lower0.9ex\hbox{$\sim$}}}  
\def\gae{\mathrel{>\kern-1.0em\lower0.9ex\hbox{$\sim$}}}  

\def\deg{\mathfont{^\circ}}  
\def\qo{\ifmmode{q_o}\else{$q_0$}\fi}  
\def\ho{\ifmmode{H_o}\else{$H_0$}\fi}   
\def\hounit{\ifmmode{{\rm km\  s}^{-1}\ {\rm Mpc}^{-        
1}}\else{${\rm    km\ s}^{-1}\ {\rm Mpc}^{-1}$}\fi}  
\def\zf{\ifmmode{z_{GF}}\else{$z_{GF}$}\fi}  
\def\kms{\ifmmode{{\rm km\ s}^{-1}}\else{${\rm km\ s}^{-1}$}\fi}

\def\msun{\ifmmode{\ {\rm M}_\odot}\else{$ {\rm M}_\odot$}\fi}  
\def\msunyr{\ifmmode{\msun \ {\rm yr}^{-1}}\else{$\msun \ {\rm 
yr}^{-1}$}\fi}

\def\sfr{\ifmmode{\dot S}\else{$\dot S$}\fi}

\slugcomment{To appear in the January 10, 2004 issue of the
  Astrophysical Journal}

\shorttitle{\chandra Observations of Abell 1068}
\shortauthors{Wise, McNamara, \& Murray}

\begin{document}

%
%
\title{The Insignificance of Global Reheating in the Abell 1068 \\
       Cluster: X-Ray Analysis}

%
%
\author{Michael W. Wise\altaffilmark{1}, 
        Brian R. McNamara\altaffilmark{2}, 
        and Steve S. Murray\altaffilmark{3}}

\affil{Massachusetts Institute of Technology, Center for Space Research \\
       Building NE80--6007, Cambridge, MA 02139--4307}

\email{wise@space.mit.edu}

\altaffiltext{1}{Massachusetts Institute of Technology, Center for Space 
                  Research, Cambridge, MA 02139--4307}
\altaffiltext{2}{Dept. of Physics \& Astronomy, Ohio University, Athens,
                 OH, 45701}
\altaffiltext{3}{Harvard-Smithsonian Center for Astrophysics, MS 70,
                 60 Garden Street, Cambridge, MA  02138}

%
%
\begin{abstract}
We report on a \chandra observation of the massive, medium redshift 
($z=0.1386$) cooling flow cluster Abell 1068.
We detect a clear temperature gradient in the X--ray emitting gas
from $kT \sim 5$ keV in the outer part of the cluster down to roughly 
2 keV in the core, and a striking increase in the metallicity of the gas
toward the cluster center. The total spectrum from the cluster 
can be fit by a cooling flow model with a total mass deposition rate
of $\sim 150 \msun \yr^{-1}$. Within the core ($r < 30$ kpc), the mass
depositon rate of $\sim 40 \msun \yr^{-1}$ is comparable to estimates 
for the star formation rate from optical data. We find an apparent
correlation between the cD galaxy's optical isophotes and enhanced
metallicity isocontours in the central $\sim 100$ kpc of the cluster.
We show that the approximate doubling of the metallicity associated
with the cD can be plausibly explained by supernova explosions
associated with the cD's ambient stellar population and the recent
starburst. Finally, we calculate the amount of heating due to thermal
conduction and show that this process is unlikely to offset 
cooling in Abell 1068.
\end{abstract}

%
%

%
%
\section{Introduction}
\label{sec:intro}

Galaxy clusters frequently have high X-ray surface brightness
cores due to thermal emission from dense gas.  Absent 
a substantial and persistent heat source, this
high density, relatively low temperature gas 
should cool on a timescale that is much less than
the age of clusters, leading to a so-called ``cooling flow''
(Fabian 1994).  The cooling rates reported from
the low resolution X-ray observatories that operated throughout the
previous two decades were often extraordinarily large, and
frequently exceeded several hundred to over two thousand
solar masses per year.  Such large cooling rates posed
a dilemma.  They imply the presence of enormous
sinks of cold gaseous and stellar matter in the cD galaxies
found at the centers of cooling flows.  Although substantial
amounts of cold gas and vigorous star formation are
now found commonly in cD galaxies centered in cooling flows
(Edge 2001, Edge et al. 2002, Jaffe \& Bremer 1997, 
Jaffe, Bremer, \& Van der Werf 2001,
Donahue et al. 2000, Falcke et al. 1998, Crawford et al. 1999,
Cardiel, Gorgas, \& Aragon-Salamanca 1998, McNamara \& O'Connell 1989),
they are found in quantities that are much smaller than
these earlier cooling rates would imply (cf., McNamara 2002).

This large discrepancy between the cooling and star formation
rates is one of the primary questions behind what has become known as 
the ``cooling flow problem''.  Simply stated, it is either a violation
of mass continuity, assuming the matter is in fact not cooling at the 
rates implied by the X--ray data, or a missing light problem, if the
matter is indeed cooling at these historically large rates but is
hidden in some exotic state.    
Recent high spatial and spectral resolution cluster observations
made with the {\it Chandra} and {\it XMM-Newton} X--ray observatories
have changed this canonical dilemma and suggested a third possibility.
The absence of detected emission lines from gas at temperatures 
less than about 3 keV in {\it XMM-Newton} RGS and \chandra HETG
observations of cluster cores implies that a large fraction of the
cooling gas does not in fact cool below X-ray emitting temperatures
(Peterson et al. 2001, Peterson et al. 2003, Wise et
al. 2004). Instead, this gas is assumed to be reheated to ambient
temperatures by one or more of several possible agents that are now
being extensively studied. These agents include mechanical and cosmic
ray reheating from supernova explosions, heat conduction from the hot
outer layers of clusters, and mechanical heating by the central radio
sources in cD galaxies. 

Images of the cooling flow regions of clusters now show a great deal
of structure, including the large cavities or bubbles in the X-ray
emission associated with the radio sources harbored by the central
cluster galaxies (e.g., McNamara et al. 2000, Fabian et al. 2000).  
The strong interactions between the radio sources generated in the
nuclei of cD galaxies and the surrounding X-ray emitting gas may
reduce the cooling rates and bring them more in line with the star
formation rates (David et al. 2001, Nulsen et al. 2002).  In addition,
{\it Chandra}'s superb spatial resolution provides the capability to
map the temperature structure of the keV gas on fine spatial scales,
which permits a direct comparison between the sites of star formation seen
at optical and ultraviolet wavelengths, and sites of rapid cooling. 
In essence, we can now reevaluate the proposition that cooling flows
are fueling star formation by comparing the cooling and star formation
rates on the same spatial scales.  In addition, the newly revealed
complexity of the gas in cluster cores has sparked a revaluation of
several possible heating mechanisms, such as AGN heating 
(Brighenti \& Mathews 2002, Ruszkowski \& Begelman 2002), heat
conduction from the hotter outer layers of clusters (Narayan \&
Medvedev 2001, Zakamska \& Narayan 2003), and supernovae.  

In this paper, we present a detailed examination of the X-ray
structure of the core of the Abell 1068 galaxy, and present some of
the most detailed temperature, and metallicity maps available for a
galaxy cluster. In a companion paper (McNamara, Wise, \& Murray 2004,
hereafter ``Paper 2''), we discuss the remarkable starburst properties
of the Abell 1068 cD galaxy and present a detailed comparison between 
the X-ray and optical properties of the cD. In these papers, we show
that 90\% of the ultraviolet light from the starburst is emerging from
the region of the cluster with the shortest cooling time, and that the
star formation rates and local cooling rates agree.  However, we stop
short of presenting this fact as proof that the starburst is being fueled
by the cooling flow, as there are independent data to suggest that the
starburst may be associated with an ongoing dynamical interaction
between the cD and a group of companion galaxies.   
In the proceeding analysis, we have adopted a redshift for Abell 1068
of $z = 0.1386$ \cite{crawford99} and a cosmology with
$\rm{H}_0$=70\hspace{.05in} 
$\rm{km\hspace{.05in}s^{-1}\hspace{.05in}Mpc^{-1}}$, 
$\Omega_M$=0.3, and $\Omega_\Lambda$=0.7.
These assumptions yield a luminosity distance of $D = 645$ Mpc,
an angular diameter distance of 505 Mpc, 
and a linear scale of 2.45 kpc per arcsec.

\section{Observations and Data Reduction}
\label{sec:data}

Abell 1068 was observed by \chandra on Feb. 4, 2001 for 30 ksec
using the ACIS-S3 back-illuminated CCD (OBSID 01652). After standard
filtering, the net exposure time was 26.8 ksec. The data were
examined for evidence of bad aspect correction and background
flares but were found to be free of both defects. 
Standard level 2 event files were created using the latest
ACIS calibration products appropriate for the observed focal
plane temperature of -120$^{\deg}$ C.
All data preparation and creation of analysis products (exposure maps,
response matrices, etc.) employed the CIAO 2.2.1 data analysis tools
and version 2.9 of the \chandra calibration database.

The X--ray morphology of Abell 1068 is complex exhibiting a highly
elliptical appearance on larger scales ($\sim 300$--400 kpc) and
complicated structure in the core ($r \lae 50$ kpc).
This morphology is illustrated in Figure~\ref{fig:flux} which shows an
exposure corrected, adaptively smoothed flux image of Abell 1068 in
the 0.3--7.0 keV band. 
The X--ray emission exhibits a strong central peak centered on
the central galaxy characteristic of ``cooling flow'' clusters.
The exact location of the X--ray peak was determined by fitting a two
dimensional, elliptical Lorentzian surface with varying minor over
major axis ratio and position angle. 
The resulting centroid position was (10 40 44.5, +39 57 12.3) which
is within 1.95 arcsec of the optical position of the central cD galaxy
(10 40 44.62  +39 57 10.2).
The best fit values for the position angle and axis ratio
were $\theta = 133^{\deg} \pm 3^{\deg}$ and $\epsilon = 0.71 \pm 0.02$,
respectively. These values were used in all subsequent spatial and
spectral analyses. 

A number of bright point sources are evident in the field of 
Abell 1068 including a bright source $\sim 23$ arcsec south
associated with the radio source B3 1037+401. Although these
sources contribute a very small fraction of the detected events
in the field (typically $\sim 100-400$ counts each), we have 
excluded them prior to extracting the spectral and surface brightness
information presented here. The CIAO tool {\tt WAVDETECT} was used
to determine positions and sizes for point sources in the field.
These sources were then excised from the event file for all
further analysis. 

All spectral analysis was done using the ISIS \cite{isis}
spectral fitting package which includes the standard XSPEC \cite{xspec}
model library as well as access to the high resolution
APED database \cite{apec}.
ACIS effective area files (ARFs) and response matrices (RMFs)
were computed using the standard CIAO tools and calibration files.
A correction for the time-dependent degradation of the ACIS
effective area due to contamination buildup on the optical 
blocking filter was also included.
All analysis was conducted using the gain--corrected, PI data.
A single RMF was computed appropriate for the centroid position
of Abell 1068. Variations in the resolution of the CCD, as captured
in the response matrices, over the region encompassing
Abell 1068 were found to be small ($\lae$10\%). Fits using RMFs
computed at other positions in the cluster as well as counts-weighted,
average RMFs gave equivalent results
within the errors. Effective area files were calculated at 
individual positions as appropriate. Finally, background files
for the spectral analysis were extracted from the edge of the
ACIS-S3 CCD, far outside the largest extracted cluster region.

\section{Spectral Analysis}
\label{sec:spectral}

In order to study the spectral character of the X--ray emission
from Abell 1068, we have performed three distinct analyses.
First, an integrated spectrum comprising the majority of the 
emission from the cluster was extracted to examine global properties
such as mean temperature, abundance, and total X-ray cooling rate 
if appropriate. 
We have also performed a spatially resolved analysis using
spectra extracted in elliptical annuli to study global trends
in the temperature, abundance, and mass deposition profile as a
function of radius in the cluster.
Finally, we have performed a fully two-dimensional mapping 
of various spectral parameters in the core of Abell 1068
to compare the complicated X--ray spatial structure with
observed structures in other wavebands.
We discuss each of these analyses and their results here in turn.

\subsection{Integrated Spectrum}
\label{sec:spect_integ}

Extracting a ``total'' spectrum for Abell 1068 is complicated by the
fact that the cluster spans more than one ACIS CCD. While variations
in the resolution from different locations on the chip are typically
small for the S3 aimpoint CCD (where the majority of the emission
from Abell 1068 is located), they can be quite sizable (factors of
$\sim$2--4 depending on energy) on front-illuminated (FI) CCDs. This
difference makes combining spectra from ACIS-S3 with spectra from
neighboring FI chips impractical. 
Consequently, we have restricted our integrated spectral analysis
to only that emission contained on ACIS-S3.
A 175$\times$129 arcsec elliptical extraction region centered on the
X--ray peak and oriented with the measured position angle was used
(see \S\ref{sec:data}).
This region corresponds to a linear extent of 0.576 Mpc (as measured
along the semi--major axis) and comprises $\sim$77\% of the total
cluster flux as determined from the surface brightness analysis
discussed below in \S\ref{sec:sb}. 
The resulting spectrum contained 65634 counts after background subtraction
and was adaptively binned to yield a minimum of 30 counts per bin
prior to analysis.

We have considered five types of models in fitting the integrated
spectrum from Abell 1068. 
In the simplest of these, the X--ray emission was represented by a
single isothermal plasma (hereafter denoted 1T) using the MEKAL model
from the XSPEC library in ISIS where the temperature and elemental
abundance were allowed to vary. 
As an intermediate case between a single temperature plasma
and a fully multi-phase model, we also considered a model with
two temperature components (2T subsequently).
In addition, three models were employed consisting of a single
temperature MEKAL model plus a multi-phase cooling flow component
(MKCFLOW in ISIS).
For all models featuring a cooling flow, the ambient temperature of
the MEKAL component and the upper bound on the cooling gas were tied.
In models with multiple components, the abundances were tied together
and allowed to vary. A foreground Galactic absorption column was
included in all five models and each model was fit twice: once
allowing the foreground Galactic absorption column to vary, and then a
second time fixing the column to the canonical value 
of $1.40 \times 10^{20}$ cm$^{-2}$. 

The lower bound on the gas temperature in the cooling flow models
was chosen to reflect differing scenarios for the fate of the cooling
material.
A scenario where the gas was assumed to fully cool below X--ray
emitting temperatures was modeled by fixing the lower temperature
bound at 0.1 keV. We have denoted this model as FC (full cooling).
However, recent XMM RGS and \chandra HETG observations of cluster
cores  have failed to detect emission lines from gas at temperatures 
less than about 3 keV indicating that cooling may be truncated
somehow (Peterson et al. 2001, Peterson et al. 2003, Wise et al. 2004).
This situation was replicated by fixing the lower temperature
bound of the cooling flow component to a value of 2 keV, consistent
with the observed radial temperature profile discussed in
\S\ref{sec:spect_spatial}, and has been denoted TC (truncated cooling).
Finally, a model where the lower limit on the temperature of the
cooling gas was allowed to be a free parameter was considered
(hereafter denoted VC for variable cooling).

The resulting spectral parameters for all 10 models are presented
in Table~\ref{tab:spectfit}.
As a general comment, all models do a reasonable job at fitting the
integrated spectrum, especially at energies above 1.0 keV,
yielding reduced $\chi^2$ values ranging from 1.1--1.3.
Differences in the fits are driven by the spectrum below 1.0 keV.
Models with freely varying absorption do a better job than models
with the canonical value. Derived columns are typically 70\% higher
than the reported value.
This difference could be related to ACIS calibration uncertainties
below 1.0 keV. Alternatively, it could represent the presence 
of excess absorption associated with the cooling material or
other cold gas in the core though this interpretation seems 
unlikely (see \S\ref{sec:spect_spatial}).
The single temperature component (1T) fits indicate a global,
mean gas temperature of $\sim$4 keV and a metallicity
of 0.5 solar abundance.
The spectral data, however, clearly indicate the presence of multiple
temperature components. This result is not surprising given the
clear temperature gradient in the spatially resolved analysis.
Formally, the best fit is achieved by the 2T model with temperature
components of $T_1 \sim 7.0$ keV and $T_2 \sim 2.0$ keV
and a reduced $\chi^2 = 1.1$. These derived temperatures are
not particularly sensitive to the value of the absorbing column.
However, this fit results in systematic residuals below 0.7 keV.
The cooling flow models give equivalent values of $\chi^2$ and
better residuals at low energies. 

\begin{figure}
\includegraphics[width=4.5in, angle=270]{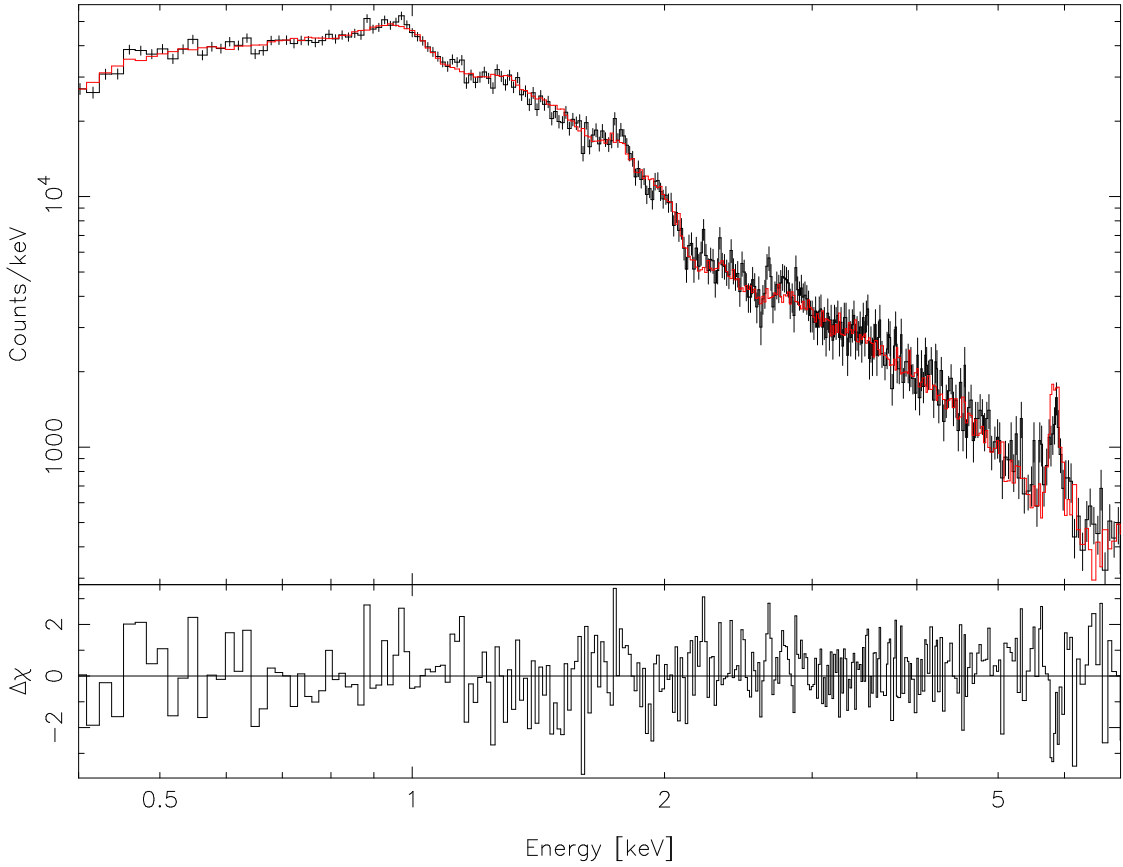}
\caption{Integrated 0.3--7.0 keV spectrum for Abell 1068 from the
175$\times$129 arcsec elliptical extraction region discussed in
the text. The spectrum has been adaptively binned to contain 
30 counts per bin. The solid line shows the best fit spectral
model, in this case the FC cooling flow model discussed in the text.}
\label{fig:spect}
\end{figure}

With CCD resolution spectra alone, it can be difficult to distinguish
between various cooling flow models. For Abell 1068, we find that the
full cooling (FC) model with free Galactic absorption produces the
lowest $\chi^2$ value and most uniformly distributed residuals
below 1.0 keV. Figure~\ref{fig:spect} shows the best fit FC model
overlaid on the data.
The resulting cooling rate for the FC model was determined to be
$\mdot_X \sim 115$---$150 \msunyr$ with an ambient gas temperature
consistent with the global mean. This rate drops to $\sim 40 \msunyr$ 
if the standard Galactic column is used. 
If the cooling is truncated at 2.0 keV, then the resulting deposition
rates are increased dramatically reaching values of 
$\mdot_X \sim 900$---$1000 \msunyr$. 
Such an increase is to be expected of course, since if the gas
cools over a smaller temperature range, then more gas must
cool in order to produce the same flux as in the FC model.
Interestingly, the variable cooling models (VC) converge to a
preferred lower temperature limit for the cooling
gas of $\sim 1.5$ keV with an accompanying small decrease in
the cooling rate relative to the TC models.
This value is consistent with the central temperature obtained
from our deprojection analysis described below (see \S\ref{sec:deproject}.
However, the additional free parameters in this model render these
fits somewhat unstable and make determining confidence limits 
difficult. So, while suggestive, one should view this result cautiously.

\subsection{Annular Spectral Analysis}
\label{sec:spect_spatial}

We have examined spatial variations in the physical state of the
X--ray emitting plasma in Abell 1068 by dividing the elliptical
region containing the previously discussed total spectrum into
a series of concentric, elliptical annuli. The annuli were 
constructed to have the same axis ratio and orientation angle
as before and were adaptively sized to include a minimum
of 5000 counts after background subtraction.
This requirement yielded 14 annuli out to a maximum, major--axis
radius of 125 arcsec. 
The spectrum outside this radius is dominated by background emission.  
Spectra were extracted from each annular region and grouped
in the same manner as the integrated spectrum.
In all subsequent discussion, quoted radii shall refer 
to distances along the cluster's major--axis except where 
specifically noted.

In fitting the spectrum from each annuli, we considered the 
various models used to describe the total, integrated spectrum
in Table~\ref{tab:spectfit}. However, the lower signal to noise
in the individual annular spectra make distinguishing between
the various models difficult. As a general result, each of the annuli
is well fit by a single temperate MEKAL plasma (the 1T model of 
Table~\ref{tab:spectfit}) with $\chi^2$ values ranging from 0.9-1.3.
Additional temperature or cooling flow components do not produce
significantly better fits. 
The exceptions to this rule are the central three annuli which
show a slight improvement in fit quality when a second temperature
component is included (2T models). 
Fits for these three annuli using the 2T models are consistent 
with cool gas at $T\sim 2.0$ keV plus a contribution from hotter
gas close to the global mean value of 4--5 keV.
As we discuss in \S\ref{sec:deproject}, the presence of these
two components is consistent with projection effects due to
hotter gas along the line of sight. Such contamination
is more easily resolved in the central annuli due to the larger
temperature difference between the intrinsic cool, core gas and
the outer, hotter cluster gas.

\begin{figure}
\includegraphics[width=4.5in, angle=90]{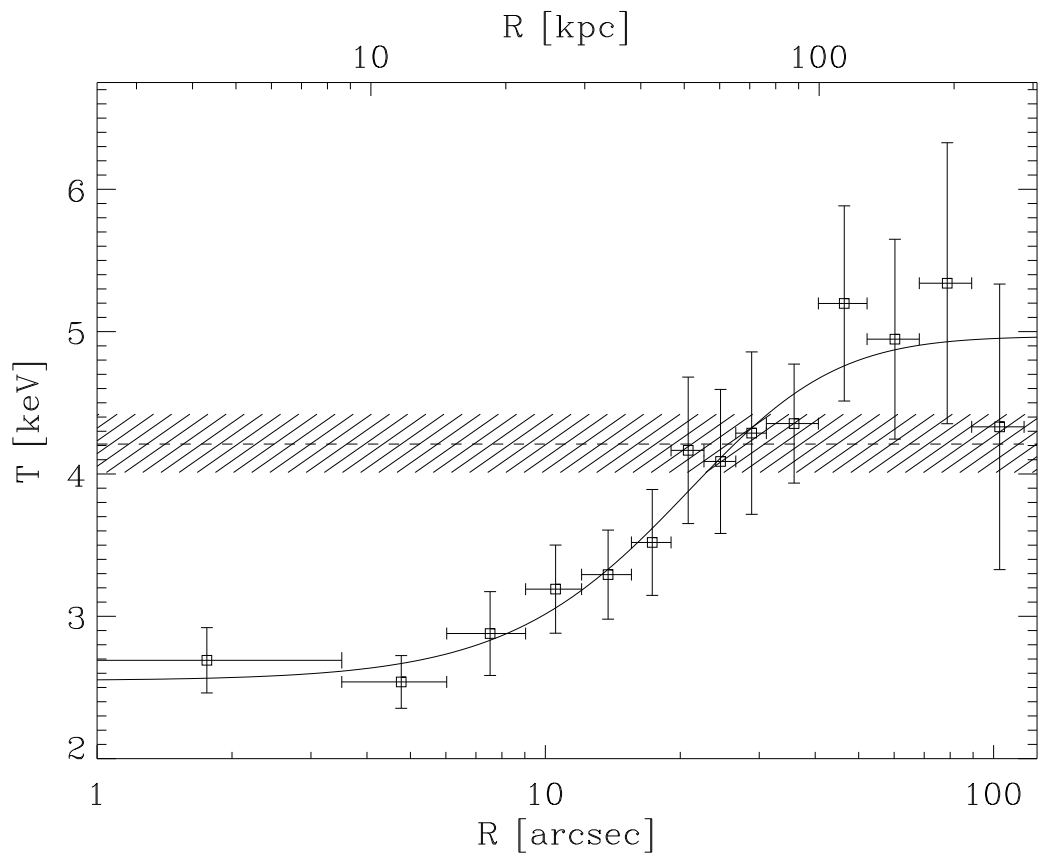}
\caption{The projected gas temperature for Abell 1068 as a function 
of projected radius along the semi--major axis in the cluster. One sigma
errors bars are indicated. 
The horizontal dashed line marks the mean gas temperature from the
best fit to the integrated spectrum shown in Figure~\ref{fig:spect},
while the hatched horizontal region depicts the 90\% confidence
interval for the temperature. The solid line shows the best fit 
analytic model.}
\label{fig:tprof}
\end{figure}

There is a clear temperature gradient from 5.5 keV in the outer
parts of the cluster down to $\sim 2.5$ keV in the core
(see Figure~\ref{fig:tprof}).
The temperature drops steadily in to a radius of $\sim 20$ kpc
($\sim 8$ arcsec) and then flattens although only two bins cover this
region. 
The innermost annulus shows a slight temperature increase
which may be due to emission from either a central AGN or the
observed starburst in Abell 1068.
Fits to this central annulus including a power--law component
to represent emission from a central AGN do not however
result in a better fit nor do they constrain the hardness of
the suspected AGN emission. For an assumed photon index of 1,
we can place an upper limit of $5.3 \times 10^{-13}$ ergs s$^{-1}$ cm$^{-2}$
on the unabsorbed, 0.3--10.0 keV flux from any central AGN.
This limit represents a maximum of $\sim 40$\% of the flux from
within the central $\sim 4$ arcsecs.
If a photon index of 2 is assumed, this limits rises to 45\%.
Given these upper limits, the observed temperature increase in the
innermost annulus is more likely to be associated with energy
injection from the starburst. 

\begin{figure}
\includegraphics[width=4.5in, angle=90]{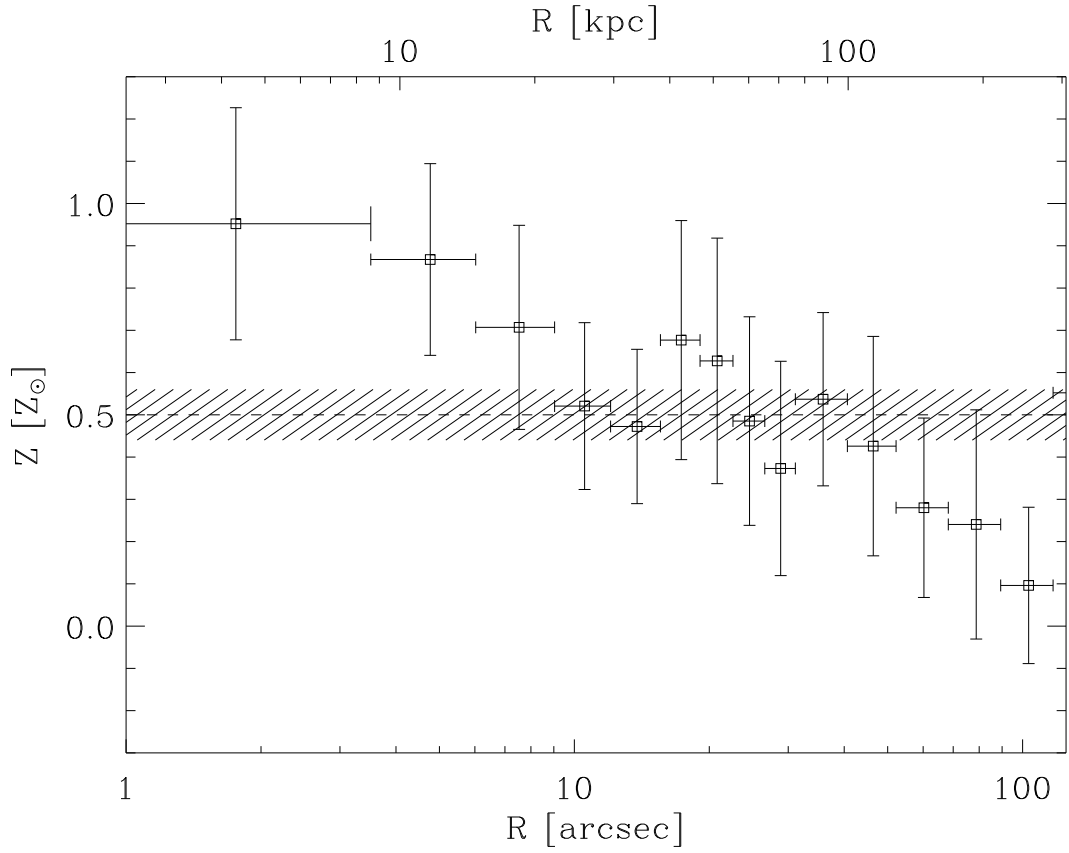}
\caption{The projected abundance profile for Abell 1068 as a function 
of projected radius along the semi--major axis in the cluster. One sigma
errors bars are indicated. 
The horizontal dashed line marks the mean abundance from the
best fit to the integrated spectrum, while the hatched horizontal
region depicts the 90\% confidence interval for the mean abundance.}
\label{fig:zprof}
\end{figure}

\begin{figure}
\includegraphics[width=4.5in, angle=90]{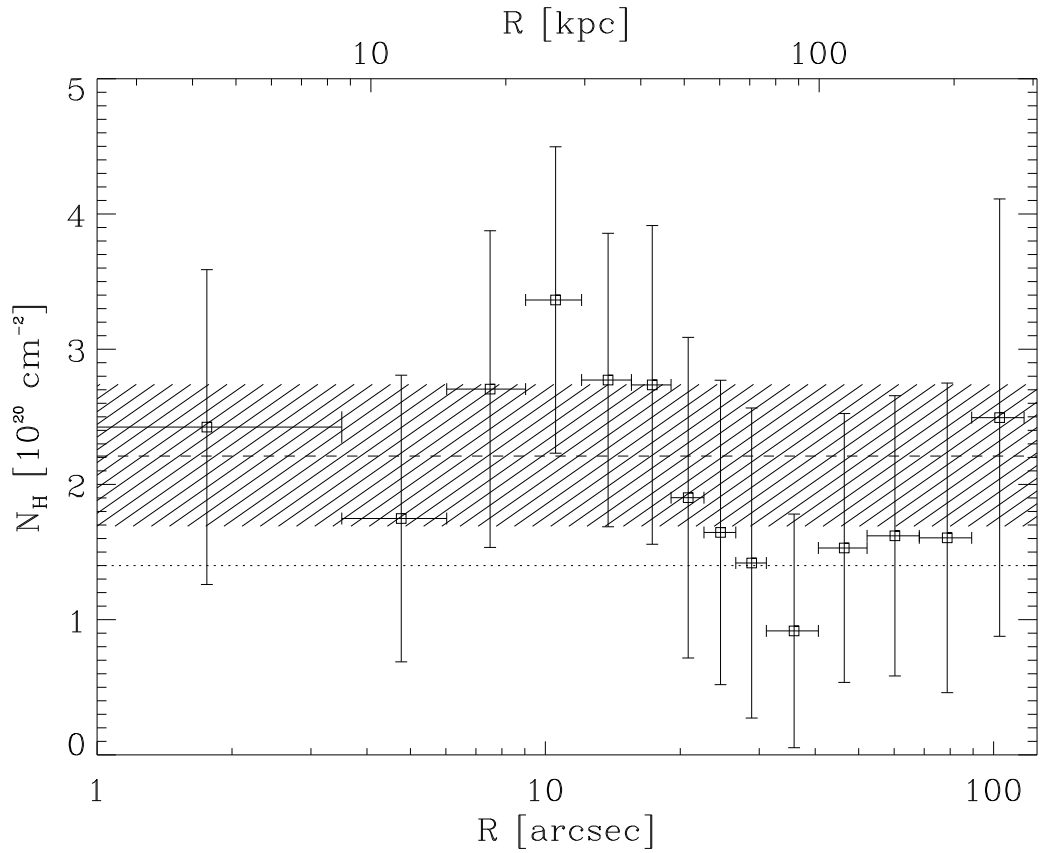}
\caption{The best fit foreground absorbing column in Abell 1068
as a function of projected radius along the semi--major axis in 
the cluster. One sigma errors bars are indicated. 
The horizontal dashed line marks the column from the
best fit to the integrated spectrum, while the hatched horizontal
region depicts the 90\% confidence interval for the mean column.}
\label{fig:aprof}
\end{figure}

The abundance in the gas shows a steady increase
toward the center of the cluster from a value
of $\sim 0.2$ solar at the outer edge to a central
value of $\sim 1.0$ solar (see Figure~\ref{fig:zprof}).
There is no evidence for a radial dependence on the amount
of absorbing column in the cluster. At all radii, the measured
column is consistent with the values obtained from fits to the
global spectrum with free absorption (see Figure~\ref{fig:aprof}).
This result would argue against the presence of intrinsic,
excess absorption associated with accumulated cooled gas
from a cooling flow.
Even if ACIS calibration uncertainties make the absolute magnitude 
of the fitted absorbing column suspect, one would expect
a relative increase toward the center of the cluster if excess
absorption associated with the cooling gas were present.

\begin{figure}
\includegraphics[width=4.5in, angle=90]{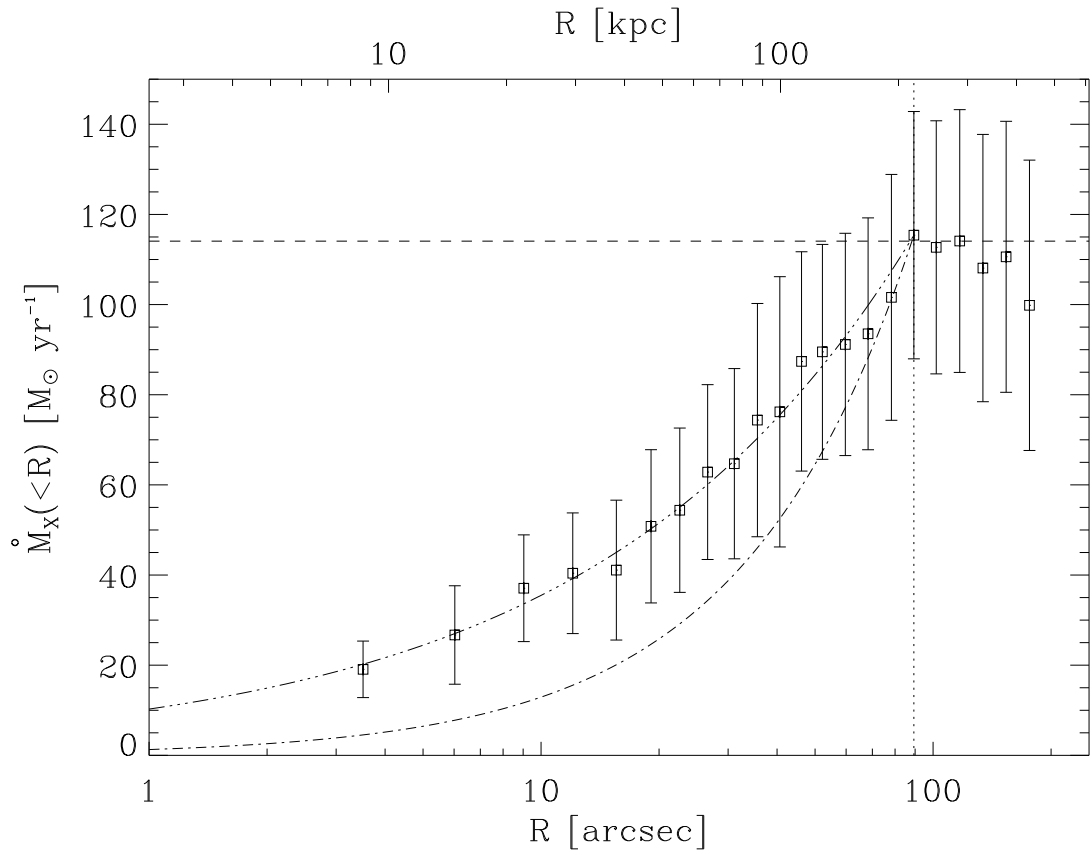}
\caption{The enclosed mass deposition rate for Abell 1068 as a 
function of projected radius along the semi--major axis in the
cluster. One sigma errors bars are indicated. The horizontal
dashed line indicates the total deposition rate from the best-fit
FC cooling flow model discussed in the text. The vertical dotted
line marks the cooling radius as derived from the temperature and
density profiles. Two analytic forms for $\mdot_X(<r)$ are depicted
by the dot-dashed and dot-dot-dot-dashed lines: the canonical 
$\mdot_X(<r) \propto r$, which does not describe the profile well,
and $\mdot_X(<r) \propto r^{0.4}$, respectively.}
\label{fig:mprof}
\end{figure}

Finally, we have measured the mass deposition rate assuming
a cooling flow model. In contrast to the annular analysis,
the cooling flow models were fit to concentric elliptical
extraction regions of increasing radius. Therefore, the measured
value of $\mdot_X(<r)$ corresponds to the total cooling rate within
a given radius. Both the TC and FC models were fit and show
similar a radial dependence qualitatively. 
Of the two cooling models, the FC model gives marginally better
fits although, again, the quality of the data does not allow
a definitive selection between the two options.
The lower signal to noise of the resolved spectra precluded fitting
the VC model. 
Figure~\ref{fig:mprof} shows the resulting profile for the
FC model. In the core region ($r \lae 25$ kpc), we find
a value of $\mdot_X(<r) \sim 40 \msunyr$. As we discuss in Paper II,
this value is consistent with the rates of massive star formation
implied by optical data in Abell 1068.
$\mdot_X$ increases steadily with radius reaching a maximum value 
of $\sim 150 \msunyr$ in the outer parts of the cluster.
Fitting a functional form to the mass deposition profile
of $\mdot_X(<r) \propto r^{\alpha}$ yields a value of 
$\alpha \sim 0.4$, somewhat flatter than the 
$\mdot_X(<r) \propto r$ relation often seen in pre--\chandra
cluster data.

\subsection{Deprojected Temperature Profiles}
\label{sec:deproject}

Due to projection effects, the temperatures measured in the previous
annular analysis actually represent the emission-weighted mean
temperature of all gas along the line of sight within a given annulus.
The spectrum extracted from a given elliptical annulus consists of a
superposition of emission from the local volume bounded by the inner
and outer ellipses and contributions from hotter gas at all larger radii.
Consequently, in a standard analysis such as in \S\ref{sec:spect_spatial},
the measured gas temperature at a given radius will be higher
than its actual local value.
To correct for the effects of this projection, we have performed
a spectral analysis on the same annular spectra used in the previous
section accounting for the contribution of hotter gas at larger radii
to the spectrum from any given annulus. 

To calculate the contributions from gas along the line of sight,
we have assumed the cluster to be composed of elliptical shells
with boundaries given by the inner and outer radii of the annuli
and an axis ratio fixed to the measured value.
Since we have no information to constrain the inclination angle 
of the cluster, we have assumed the orientation to be perpendicular
to the line of sight although the equations for the volume
contributions are equivalent for the edge-on case.
The intrinsic spectrum of each volume shell was modeled as
a single MEKAL plasma with variable normalization, temperature,
and elemental abundance and the fractional contribution of each
volume shell to the various annuli was then calculated.
A foreground Galactic absorption component was included and
allowed to vary.
The spectra from all 14 annuli were then fit simultaneously
to determine the intrinsic temperature profile in the cluster.
This entire procedure has been implemented as a custom model
within ISIS \citep{isis} and is described in detail in
Wise \& Houck (2004). 
Similar techniques have been used by other groups to correct for
projection in clusters such as Abell 2390 (Allen, Ettori, \& Fabian 2001)
and Abell 2199 \cite{johnstone02}. 

\begin{figure}
\includegraphics[width=4.5in, angle=90]{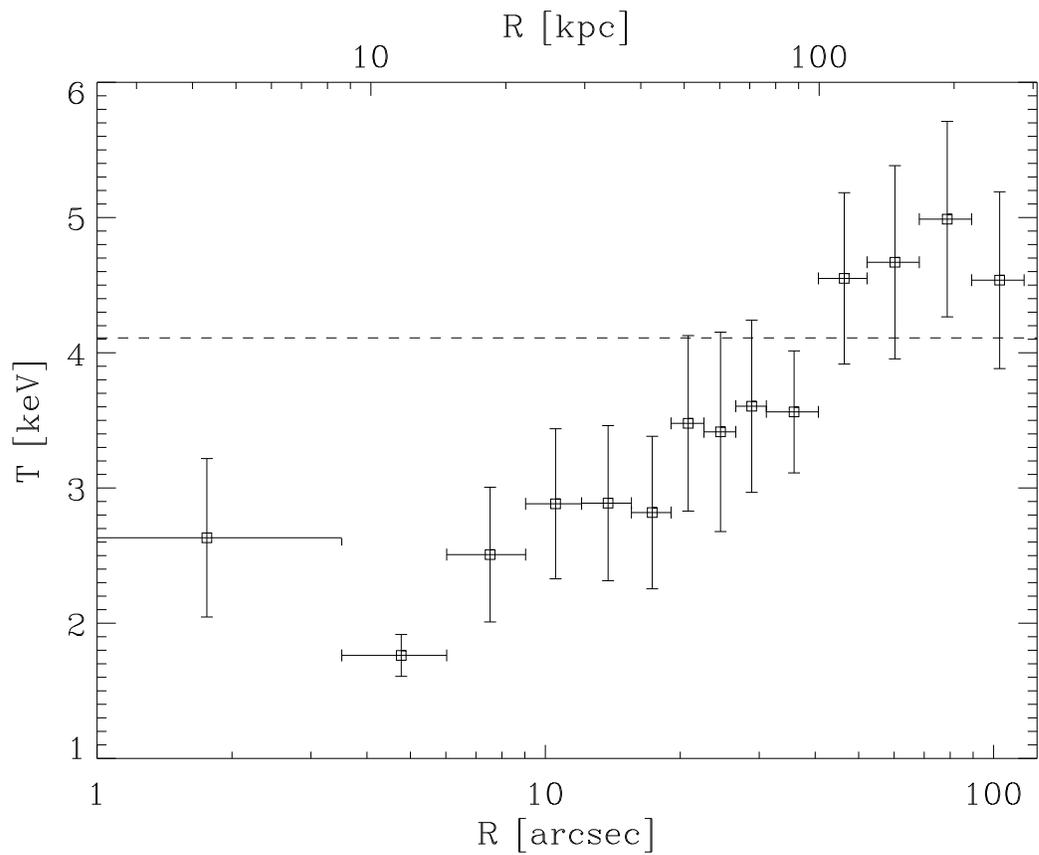}
\caption{The deprojected gas temperature for Abell 1068 as a function 
of projected radius along the semi--major axis in the cluster. One sigma
errors bars are indicated. The horizontal dashed line marks the mean
gas temperature from the best fit to the integrated spectrum shown in
Figure~\ref{fig:spect}.} 
\label{fig:dtprof}
\end{figure}

Qualitatively, the deprojected temperature profile is quite
similar to that obtained from the standard annular analysis
(see Figure~\ref{fig:dtprof}), although with somewhat larger
errors. At large radii, the temperature agrees with the simple
analysis as one would expect since contamination from other shells  
drops with increasing radius. The biggest change occurs for
the central temperature where we obtain obtain a somewhat cooler
minimum temperature of $1.8 \pm 0.2$ keV as compared to the 2.5 keV
obtained in the analysis of \S\ref{sec:spect_spatial}. 
As noted in the simple annular analysis, the innermost annulus
remains somewhat hotter than the minimum temperature possibly
due to contamination from an unresolved, central AGN.
Values for the abundance were consistent with the simple
annular analysis and the Galactic column was determined to be
2.9 $\times{10}^{20}\rm{{cm}^{-2}}$.

\subsection{Temperature Map}
\label{sec:tmap}

\begin{figure}
\includegraphics[width=4.9in, angle=90]{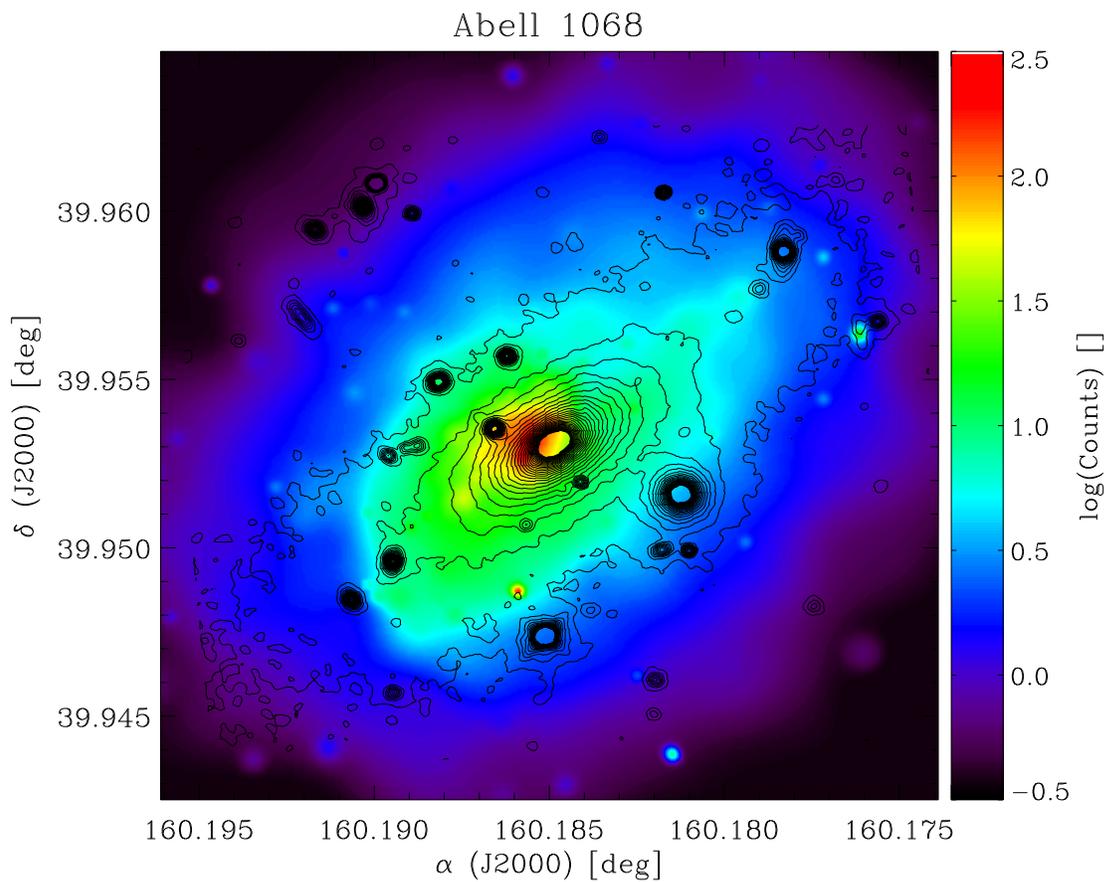}
\caption{An exposure corrected, 0.3-7.0 keV image of the central
$80^{\arcsec}\times80^{\arcsec}$ of Abell 1068 from the 26.8 ksec
ACIS-S exposure. The image has been adaptively smoothed using the
CIAO tool {\tt CSMOOTH}. The optical contours corresponding to the
HST R--band (F606W) surface brightness are also shown.
Note the offset between the X--ray centroid and the optical center of the
cD galaxy.}
\label{fig:flux}
\end{figure}

It is apparent even by visual inspection that the X--ray emission
from the core of Abell 1068 exhibits a complicated morphology
(see Figure~\ref{fig:flux}). As we discuss in more detail in
Paper II, the optical data show similar complicated structures
including condensed knots of extremely blue light presumably
associated with sites of ongoing star formation 
(McNamara, Wise, \& Murray 2004).
In order to examine possible correlations between the plasma
properties and features seen in other wavebands, we have constructed
2D maps of the gas temperature and abundance in Abell 1068.
These maps were computed using a custom module within the ISIS
\citep{isis} spectral analysis package 
which automates the process of extracting spectra, generating
appropriate ARFs and RMFs, and then fitting a given spectral model 
(see \cite{tmap} for a detailed description).
 
The maps were generated using a grid of adaptively sized 
extraction regions selected to contain a minimum of 1000 counts
in the 0.5--7.0 keV band. The grid was selected to span a 
$300 \times 300$ arcsec region centered on Abell 1068. Due to the
drop in surface brightness with increasing radius, the size of the
extraction regions increases with radius. Consequently, sharp spatial
features are not well resolved in the outer parts of the maps.
In the core ($r \lae 40$ kpc), however, the extraction regions 
range in size from $\sim 2$--10 arcsecs and features in the
temperature or abundance distributions on these scales can be
identified. Within each extraction region, the resulting spectrum was
fit with a MEKAL plasma model including foreground Galactic absorption
fixed at the nominal value of 2.41 $\times{10}^{21}\rm{{cm}^{-2}}$ as  
determined from the integrated fits in \S\ref{sec:spect_integ}.
The plasma temperature and abundance were allowed to vary and 
single-parameter, 90\% confidence limits were computed for both
parameters at each map point. Typical errors for the measured 
temperatures and abundances range from 10\%---30\% over the map with
the most accurate measurements corresponding to the central--most regions
where the surface brightness is highest. Fits allowing the Galactic
column to vary produced similar maps within the measured uncertainties
and in any event were morphologically equivalent. 

\begin{figure}
\includegraphics[width=4.9in, angle=90]{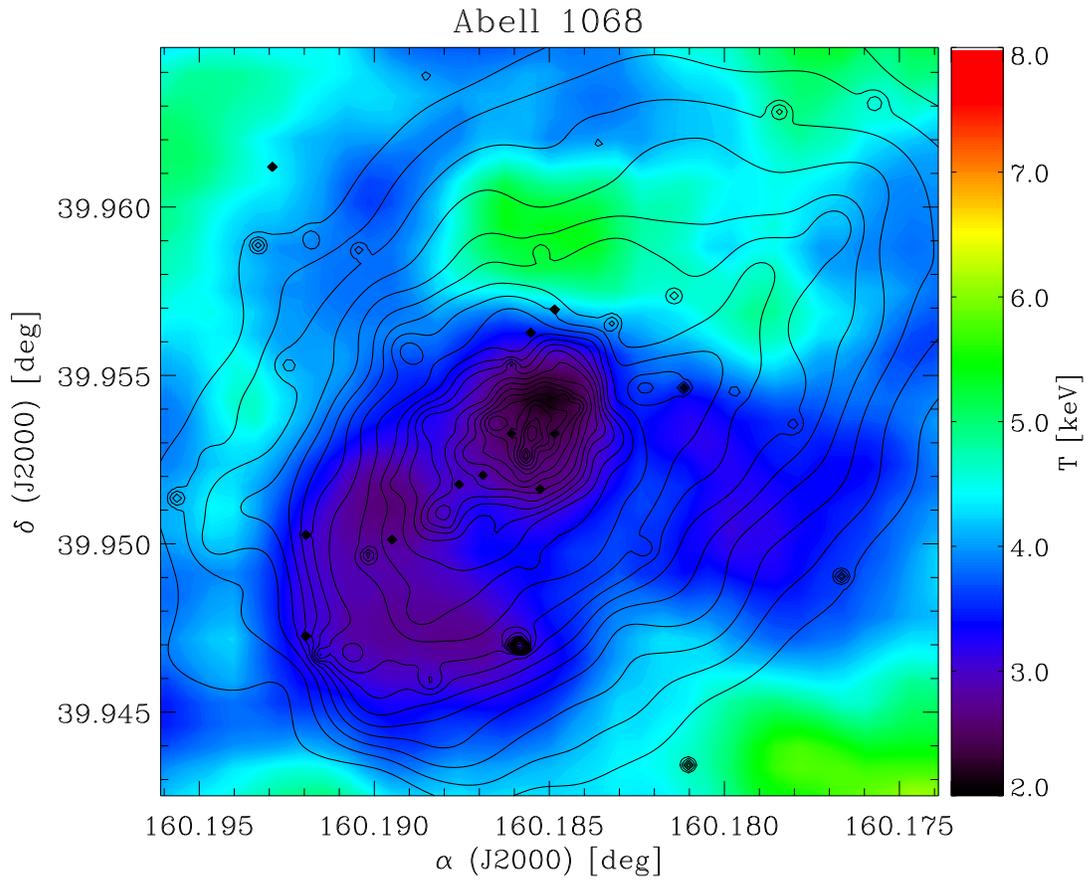}
\caption{A map of the gas temperature for the central
$80^{\arcsec}\times80^{\arcsec}$ in Abell 1068.
The overlaid contours show the surface brightness from the 
adaptively smoothed flux image shown in Figure~\ref{fig:flux}.
The color bar on the right indicates the temperature scale in keV.
Errors for the measured temperatures range from 10\%---30\%
over the map.}
\label{fig:tmap}
\end{figure}

The temperature map for the central $80^{\arcsec}\times80^{\arcsec}$
of Abell 1068 is shown in Figure~\ref{fig:tmap}.
The 0.3-7.0 keV X--ray surface brightness contours are overlaid
for comparison.
Although the overall temperature structure is consistent with the
radial temperature profile presented in Figure~\ref{fig:tprof},
significant asymmetry in the gas temperature is apparent.
To the southeast of the cD position, an arc of 2--3 keV is present.
This extension of cooler gas is spatially coincident ($r \sim 20$ arcsec)
with several companion galaxies and may represent either gas
associated with the ISM in these galaxies or stripped galactic gas
from these companions.
The region of coolest emission ($T \sim 2.5$ keV) corresponds
to the peak of the X--ray surface brightness profile and is
offset from the optical centroid of the cD galaxy by $\sim 2$ arcsec.
However, this scale is comparable to the map resolution in the core
and so may be consistent with a zero offset. 
As we discuss in Paper II, the sites of coolest
X--ray emission in Abell 1068 correspond to over 90\% of the
emergent optical blue light from the starburst. We defer a detailed
discussion of the implications for the star formation history
in Abell 1068 to that work (McNamara, Wise, \& Murray 2004).

In contrast, to the northwest, the temperature rises relatively steeply
as one moves away from the central cD increasing from 2.5 keV to
5 keV within $\sim 10$ arcsec ($\sim 25$ kpc). This rapid rise
in temperature corresponds to a sharp decline in the X--ray 
surface brightness at this position as seen in Figure~\ref{fig:flux}.
This position is also coincident with an inflection point in the 
azimuthally averaged surface brightness profile discussed in \S\ref{sec:sb}.
Such temperature jumps and surface brightness discontinuities are
characteristic of the ``cold fronts'' observed in many other clusters
and may represent evidence for previous merger activity in Abell 1068
(Markevitch et al. 2000; Vikhlinin, Markevitch, \& Murray 2001).
We note however that the surface brightness inflection observed in
Abell 1068 is much weaker than that typically seen in clusters
featuring cold fronts such as Abell 2142 (Markevitch et al. 2000).
We defer a more detailed analysis of this feature to a later work.

\subsection{Abundance Map}
\label{sec:amap}

\begin{figure}
\includegraphics[width=4.9in, angle=90]{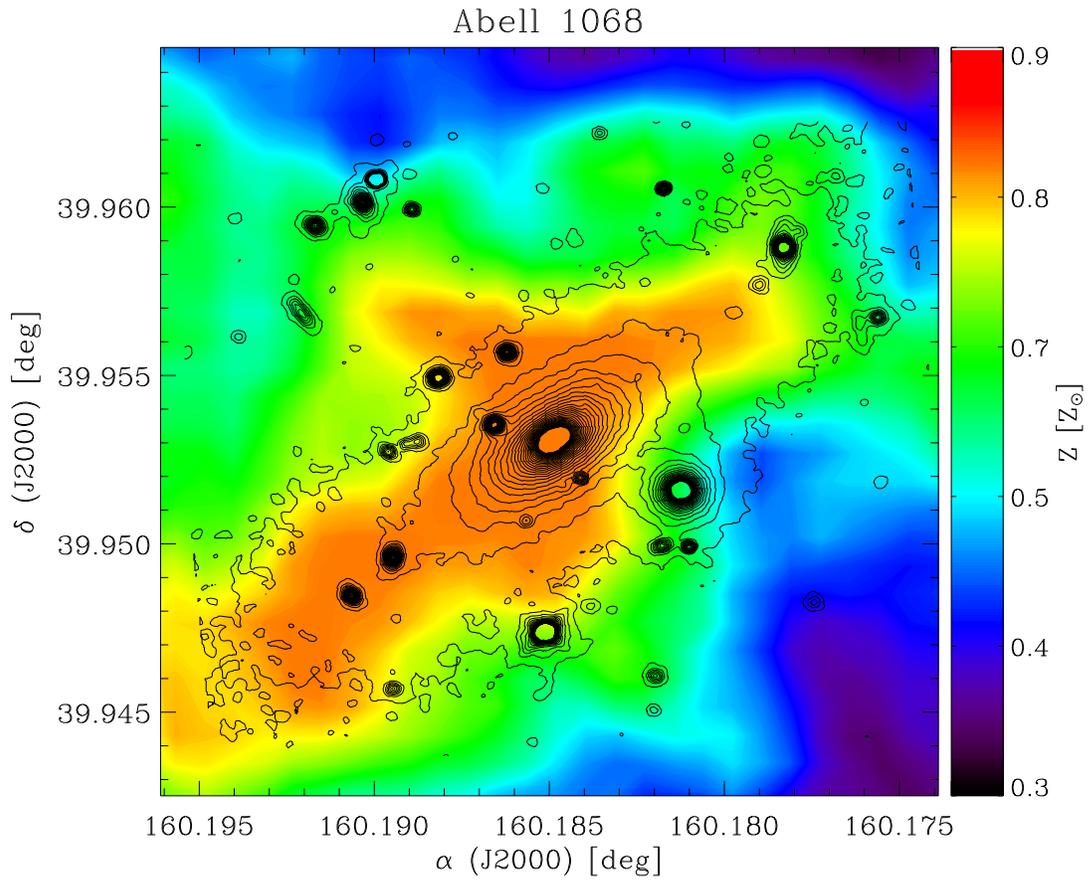}
\caption{A map of the elemental abundance for the central
$80^{\arcsec}\times80^{\arcsec}$ in Abell 1068.
The optical contours corresponding to the
HST R--band (F606W) surface brightness are shown.
The color bar on the right indicates the abundance 
in units of solar abundance.
Errors for the measured abundances range from 10\%---30\%
over the map.}
\label{fig:amap}
\end{figure}

The morphology of the abundance map in Abell 1068 is even more
striking. Figure~\ref{fig:amap} shows the metallicity as determined
from the MEKAL fits for the same $80^{\arcsec}\times80^{\arcsec}$
central region. 
An optical image of the R--band light (F606W filter) in Abell 1068 was
obtained from the HST archive and is overlaid for comparison.
There is a clear pattern of enrichment extending
along the semi--major axis of the cluster and the central cD galaxy.
Several filamentary extensions of roughly solar metallicity gas are
also apparent and in all cases extend from the center to the
position of the various companion galaxies in the cluster.
Since the map abundances are measured in projection, local
abundances may be somewhat higher.

Several researchers have found a trend between increasing
metal abundance and the presence of a cD galaxy in the
central regions of clusters (Fukazawa et al. 2000, Irwin \& Bregman 2001,
David et al. 2002).  This trend may be somewhat stronger in
cooling flow clusters (Irwin \& Bregman 2001).  Two of these studies 
were carried out using the low resolution {\it ASCA} and 
{\it BeppoSAX} observatories, and were based on radially-averaged
trends between metallicity and radius.  These trends suggest
that the correlation between the presence of the cD and 
the central metallicity enhancement came about primarily as 
a result of type one supernovae (SNe Ia) in the cD's stellar populations,
with some contribution of type two supernovae (SNe II) from star formation
in the cooling flows. 

In Figure~\ref{fig:amap}, we show that there is
a remarkable correlation between the location and shape of 
the Abell 1068 cD galaxy's 
optical isophotes and the isometallicity contours.
The strongly flattened appearance of both the isophotes and
the metallicity contours renders a chance coincidence unlikely.
It suggests strongly that the presence of the cD and regions
of enhanced metallicity are strongly coupled.  
The average cluster metallicity beyond the cD is
$Z_0\sim 0.4$, in solar units, and rises to $Z_{\rm cD}\sim 0.8$
at the location of the cD. This metallicity increase has likely occurred
either by the external introduction of enriched gas during a group merger
or by an internal process related to mass loss from the stars in
the cD or the cluster at large.  Given the correlation between
the cD's optical isophotes and the contours of highest metallicity,
we will examine whether mass loss from the general population
of stars in the cD and the ongoing star formation would
be capable of producing the observed metallicity enhancement.

We can characterize the metallicity of the keV gas in the vicinity
of the cD as $Z_{\rm cD}=Z_0 + \delta Z$, where $\delta Z$ represents
the metal enhancement due to supernova explosions, then
\begin{equation}
Z_{\rm cD}\simeq Z_0 + \Bigl({y_I + y_{II} \over M_H }\Bigr)\times \Bigl({M_Z 
\over M_H }\Bigr)^{-1}_\odot.
\label{eqn:starform}
\end{equation}
In order to calculate the supernova type Ia rate, we adopt Tammann's (1974) 
value $4\times 10^{-13} L_{\rm B}~{\rm yr}^{-1}$.  The integrated
$R$-band absolute magnitude of the cD out to a 170 kpc radius
is $-24.0$, which corresponds to $L(R)=2\times 10^{11}L_\odot$.
Adopting an average color for a typical cD of $(B-R)=1.6$ (Peletier
et al. 1990), we find an integrated $B$-band luminosity of
$L(B)=1.4\times 10^{11}L_\odot$.  This corresponds to a SNe Ia
rate of $0.03 ~{\rm yr}^{-1}$. For the the iron yield per SNe Ia
we adopt $0.74 \msun /{\rm SNe~ I}$ (cf. David et al. 2001),
giving a total yield of $y_I\sim 2\times 10^7\msun (t/{\rm Gyr})$.

\begin{figure}
\includegraphics[width=4.5in, angle=90]{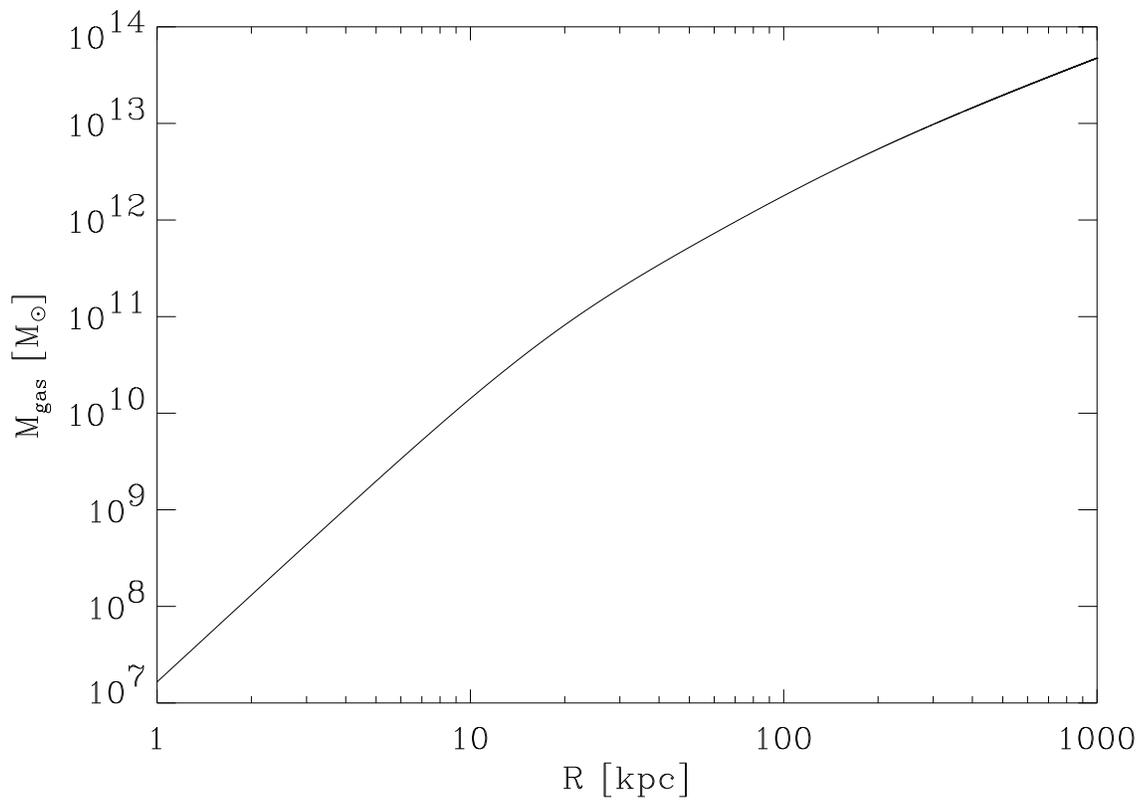}
\caption{Gas mass profile for Abell 1068.}
\label{fig:mass}
\end{figure}

In addition to the yields from SNe Ia, there will be a significant
contribution of metals from SNe II associated with the
ongoing star formation in the cD.  Assuming an iron
yield of $0.12 \msun  /{\rm SNe II}$, and a supernova rate
of 1 per $100 \msun$ of star formation (cf. David et al. 2001), we obtain a
total yield of $y_{II}\sim 1.2\times 10^{-3}M_{\rm AP}~\msun$.
Adopting the luminosity masses for the accretion populations
in Table 2, we find $y_{II}\sim 8\times 10^5-4\times 10^6 \msun$.
In order to match the observational constraints, the supernova
yields must approximately double the ambient metallicity of
$\sim 0.4$ solar by seeding the ambient gas with metals.
>From Figure~\ref{fig:mass}, we find an  integrated gas mass within a 170 kpc 
radius of $3.4\times 10^{12}\msun$.  Further adopting a solar iron abundance 
$(M_Z/M_H)\odot =4.67\times 10^{-5}$, we find that the 
observed central abundance gradient would accrue after only
$\sim 3$ Gyr.  This conclusion is largely independent of the
SNe II yields from  the starburst.  Assuming SNe II yields from
the maximum mass of the accretion population permitted by the data,
$M_{\rm AP}= 3.1\times 10^9 \msun$ 
(cf. Table 2), the mass ratio of SNe Ia to SNe II yields is 15.
The implications of this result for the starburst in Abell 1068
is discussed in more detail in Paper II.

To summarize, the metallicity gradient associated with the
cD galaxy can be almost entirely accounted for
by type Ia supernovae originating in 
the ambient stellar population of the cD over a reasonable 
age for the galaxy.  However, in the
central starburst region, the 
SNe II yields from the starburst will become increasingly important.
Furthermore, the central metallicity will almost certainly continue to grow
as the roughly $4\times 10^{10}\msun$ of remaining molecular
fuel (Edge 2001) is consumed by star formation.  We should
emphasize the factors of several uncertainty in both
the star formation masses, supernova rates, and yields.
However, the numbers taken at face value, coupled with the
observed correlation between the optical and metallicity
isocontours, present persuasive case for 
cD galaxy-driven metallicity gradients in the keV gas.

\section{Surface Brightness Profiles}
\label{sec:sb}

Despite the apparent spatial complexity evident in the core of 
Abell 1068, the overall surface brightness in the cluster
can still be well described by an analytic $\beta$ model profile.
This profile combined with the temperature information from 
\S\ref{sec:spect_spatial} can be used to determine the density profile, 
and various other physical properties of the ICM.
Surface brightness profiles were calculated by constructing
image mosaics of all chips in the observation in three bands:
a total band image from 0.3-7.0 keV, and soft and hard bands 
spanning 0.3-2.0 keV and 2.0-7.0 keV, respectively. 
Point sources were filtered out as discussed in \S\ref{sec:data}.
Appropriately weighted exposure maps were then created for each band
in order to accurately correct for variations in the detector
efficiency over the field.
Finally, the surface brightness profiles were calculated in elliptical
annuli, 1 arcsec in width with a position angle and axis ratio
as discussed in \S\ref{sec:data}. The profiles were determined out
to a maximum radius of 16.6 arcminutes or 2.8 Mpc.

\begin{figure}
\includegraphics[width=4.5in, angle=270]{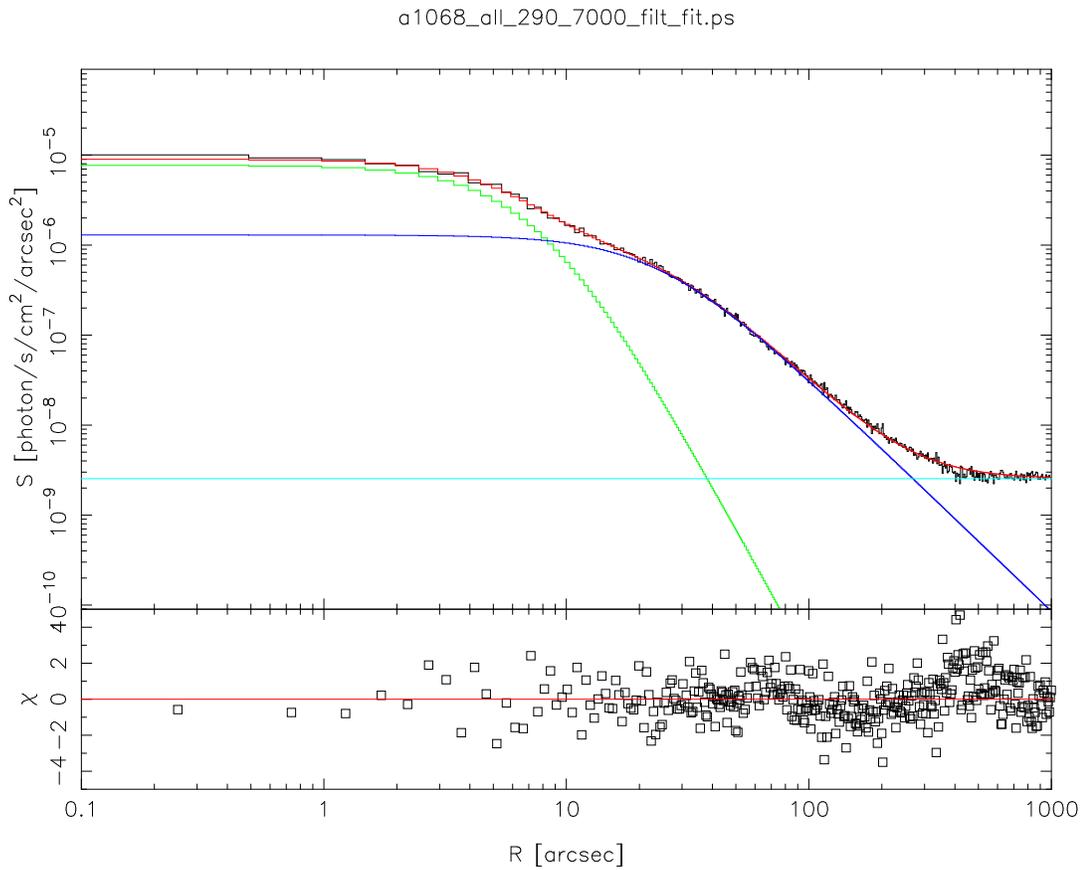}
\caption{Radial surface brightness for Abell 1068 in the 0.3-7.0 keV
band accumulated in 1 arcsec annular 
bins. The solid lines indicate the two components of the best fit
double $\beta$ model discussed in the text. The horizontal dotted line
represents the best fitting background value. The lower pane shows the
residual deviations from the fit.}
\label{fig:betafit}
\end{figure}

Figure~\ref{fig:betafit} shows the resulting surface brightness
profile in the 0.3--7.0 keV band. Profiles for the soft and hard
bands look similar.
An inflection in the profile is apparent at a radius 
of $\sim 10$ arcsec ($\sim 25$ kpc).
Attempts to fit the surface brightness profiles with single
$\beta$ models yielded unacceptable fits.
Consequently, we have used a double $\beta$ model of the form
\begin{equation}
I(r) = I_B +
       I_{1} \left( 1 + {r^2 \over r_{1}^2} \right)^{-3\beta_1+\frac{1}{2}} + 
       I_{2} \left( 1 + {r^2 \over r_{2}^2} \right)^{-3\beta_2+\frac{1}{2}}
\label{eqn:dbeta}
\end{equation}
to describe the surface brightness profiles.
Each component has fit parameters $(I_i, r_i, \beta_i)$, where
the normalizations and core radii are denoted by $I_i$ and $r_i$,
respectively. The term $I_B$ is a constant representing the 
contribution of the background.
A detailed discussion of the double $\beta$ model and its advantages
for representing cluster X--ray surface brightness profiles can
be found in Xue \& Wu (2000) and Ettori (2000).

\begin{figure}
\includegraphics[width=4.5in, angle=270]{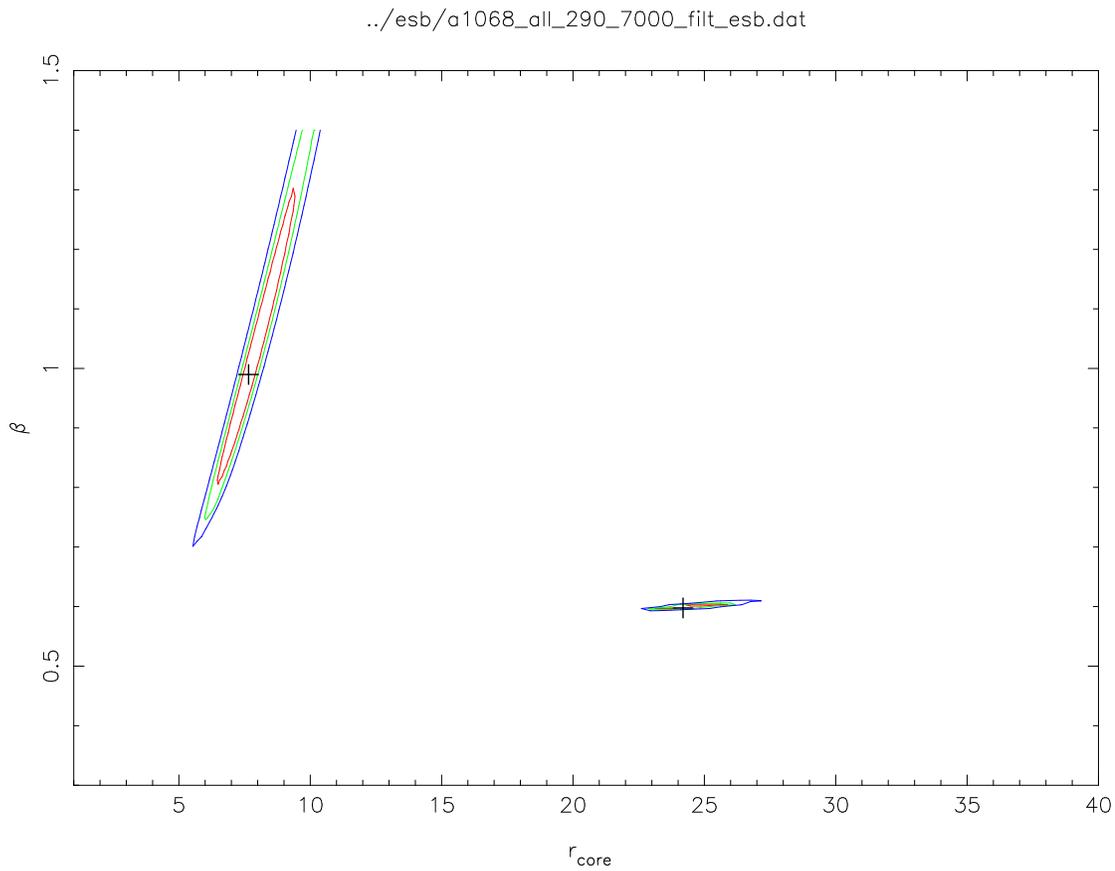}
\caption{Confidence contours for the two components of the double $\beta$
model fit to the surface brightness profile shown in Figure~\ref{fig:betafit}. 
Contours are shown for 1$\sigma$, 2$\sigma$, and 3$\sigma$. The
crosses indicate the location of the best fit ($r_c$, $\beta$) values
for the two components.}
\label{fig:betaconf}
\end{figure}

Table~\ref{tab:betafit} shows the results from fitting 
equation~\ref{eqn:dbeta} to the profiles for the total, soft,
and hard bands. In all three bands, a broad component corresponding
to the outer cluster gas is present with fit values of $r_c \sim 24$
arcsecs for the core radius and $\beta \sim 0.59$, consistent within
the measurement errors. For the total 0.3-7.0 keV, the fit to the
inner component yielded values of $r_c \sim 8$ arcsecs and
$\beta \sim 0.99$. As Figure~\ref{fig:betaconf} illustrates,
these two components are distinct at the 99\% confidence level.
Comparing the soft and hard band fits, we find that the inner $\beta$
component in the soft, 0.3-2.0 keV band is more extended relative
to the hard band with fit parameters of $r_c \sim 11$ arcsec and
$\beta \sim 1.0$. This extended soft core is consistent with the
flattening of the temperature profile inside 10 arcsec discussed in
\S\ref{sec:spect_spatial} and shown in Figure~\ref{fig:tprof}.
>From 2.0-7.0 keV, the best fit value of the core radius for the
inner component is $r_c \sim 4$ arcsecs, possibly due to the presence
of non--thermal emission from the unresolved, central AGN
or heating from the starburst. 
Reduced $\chi^2$ values for the total, soft, and hard band
fits were 1.8, 1.2, and 1.2, respectively.

We note that recent models of bubbles in cluster atmospheres
associated with central AGN can produce ``kinks'' or inflections
in the azimuthally averaged surface brightness similar to that
seen in Figure~\ref{fig:betafit}.
However, no other obvious evidence for bubbles is apparent
in the images of Abell 1068. Also, the strong correlation between
the optical isophotes and the abundance map presented in 
Figure~\ref{fig:amap} is difficult to reconcile with significant
mixing of the ICM due to bubbles.

\section{Physical Conditions in A1068}
\label{sec:physics}

\subsection{Density Profile}
\label{sec:dens}

Following the derivation of Xue \& Wu (2000), we assume that the two
$\beta$ model components in equation~\ref{eqn:dbeta} correspond to
two phases of gas in the ICM with potentially different electron
temperatures $T_1$ and $T_2$. With this assumption,
equation~\ref{eqn:dbeta} can be inverted to yield the electron
number densities for the two components as well as the combined
electron number density, $n_e(r)$,
\begin{equation}
n_e(r) = \sum_i n_{ei}(r) = \left[ n_e(0) \sum_i \tilde{n}_{ei}(r) 
          \right]^{1/2},  ~~~~i=1,2 
\end{equation}
\begin{equation}
n_{ei}(r) = \left[ \frac{n_e(0)}{n_e(r)} \right] ~\tilde{n}_{ei} (r)
\end{equation}
\begin{equation}
\tilde{n}_{ei}(r) = n_{ei}(0) 
\left( 1 + {r^2 \over r_{i}^2} \right)^{-3\beta_i}
\end{equation}
where the core radii, $r_i$, and exponents, $\beta_i$, are the same 
as before and $n_e(0)$ represents the central, total electron density.
The central number densities for the individual components are related
to the fitted surface brightness parameters, $(I_i, r_i, \beta_i)$, by
\begin{equation}
n^2_{ei}(0) = \left[ \frac{4\pi^{1/2}}{\alpha(T_i) g_i \mu_e} \right]
              ~\left[ \frac{\Gamma(3\beta_i)}{\Gamma(3\beta_i - 1/2)}
               \right]
              ~\left( \frac{I_i}{r_i} \right) ~A_{ij}
\label{eqn:cdens1}
\end{equation}
where
\begin{eqnarray}
\label{eqn:cdens2}
\frac{1}{A_{ij}} & = 1 + \left( \frac{g_i}{g_j} \right)
                   \left( \frac{r_i I_i}{r_j I_j} \right)
                   \left( \frac{T_i}{T_j} \right)^{1/2}
                  ~\left[ \frac{\Gamma(3\beta_j) ~\Gamma(3\beta_i - 1/2)}
                    {\Gamma(3\beta_i) ~\Gamma(3\beta_j - 1/2)} 
                   \right] , \\ 
 ~~~~ & j=1,2 ~~\mbox{and} ~~j \neq i \nonumber
\end{eqnarray}
and $g_i$ is the Gaunt factor for component $i$ 
and $\alpha(T)$ is the emissivity due to thermal bremsstrahlung
as given in Xue \& Wu (2000).

\begin{figure}
\includegraphics[width=5.5in, angle=0]{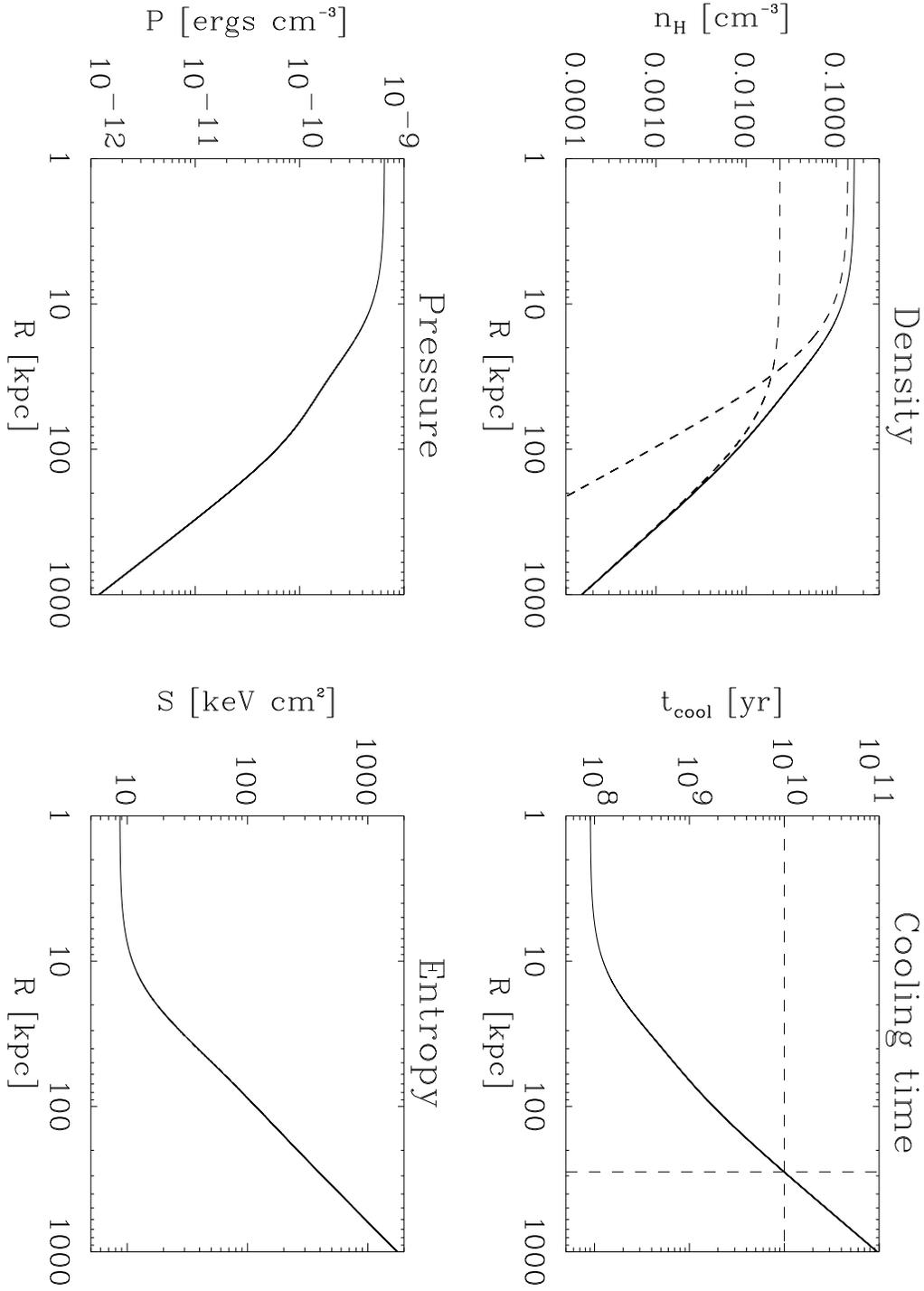}
\caption{Radial variation of (a) electron density, (b) isobaric
cooling time, (c) pressure, and (d) entropy. In panel (a), the
dashed lines depict the two components of the electron density
from the double $\beta$ model fit to the surface brightness
distribution. In panel (b), the horizontal line indicates a
cooling time equal to $10^{10}$ years and the vertical line
indicates the corresponding cooling radius.}
\label{fig:dens}
\end{figure}

The choice of values for $T_1$ and $T_2$ is somewhat uncertain;
however, as equations~\ref{eqn:cdens1} and \ref{eqn:cdens2} indicate,
the temperature dependence of the central electron densities is
fairly weak. We have chosen the minimum and maximum temperature
values from the deprojected temperature profile 
discussed in \S\ref{sec:deproject} for the narrow and extended
components, respectively. Using $T_1 = 1.76$ keV and $T_2 = 5.0$ keV,
we obtain central densities of ~$n_{e1}(0) = 0.135$ cm$^{-3}$ and
~$n_{e2}(0) = 0.024$ cm$^{-3}$. The resulting density profiles are
show in Figure~\ref{fig:dens}. We note that using the two temperatures
derived from the two-phase model (2T) fits in Table~\ref{tab:spectfit}
resulted in central densities within 5\% of these values.
Of course, if the gas is truly multiphase as the spectral analysis in
\S\ref{sec:spect_integ} indicates, then representing the temperature
structure using only two components is simplistic. However, the 
observed surface brightness profile does not allow additional
components to be unambiguously identified.

Using the derived density profiles, the spectral fitting results
of Table~\ref{tab:spectfit}, and the assumption of hydrostatic
equilibrium, we have computed various estimates for the gas
and total mass in Abell 1068. 
Table~\ref{tab:mass} gives the gas mass and total mass within 0.5 Mpc
for the various spectral models discussed in \S\ref{sec:spect_integ}
following the method discussed in Hicks et al. (2002).
Figure~\ref{fig:mass} shows the gas mass profile for the best fit
FC model.
In addition to the total cluster estimates, we have calculated 
the mass associated with the central, condensed $\beta$ model
component separately. These estimates are labeled ``Core'' in
Table~\ref{tab:mass} and assume a central temperature of 
1.76 keV taken from the deprojected temperature profile
discussed in \S\ref{sec:deproject}.

\subsection{Cooling Time}
\label{sec:cool}

The characteristic time that it takes a plasma to cool isobarically
through an increment of temperature $\delta$T can be written
\begin{equation}
t_{cool} = \frac{5}{2}~\frac{k ~\delta T}{n_e \Lambda(T)} 
\label{eqn:tcool}
\end{equation}
where $n_e$ is the electron density, $\Lambda$(T) is the
total emissivity of the plasma (the cooling function), and k is
Boltzmann's constant.  For isochoric cooling, 5/2 is
replaced by 3/2. 
Utilizing the temperature profile from \S\ref{sec:spect_spatial}
and the density profile from \S\ref{sec:dens}, we can calculate
the cooling time as a function of radius in the cluster.
Figure~\ref{fig:dens} shows the resulting cooling time as a 
function of radius in Abell 1068. The central cooling time in the
core reaches a minimum value of $9 \times 10^7$ yrs.
The ``cooling radius'' where the cooling time exceeds the age of the
cluster, canonically taken to be $10^{10}$ yrs, is 282 kpc.
As we show in Paper II, 98\% of the blue light associated with
star formation in Abell 1068 occurs inside the central $\sim 40$ kpc
where the cooling time is $\leq 5 \times 10^8$ yrs.

\subsection{Entropy}
\label{sec:entropy}

Similarly, we can use the density and temperature information to 
calculate the entropy in the cluster gas as a function of position
using
\begin{equation}
S = T / n^{2/3}
\label{eqn:entropy}
\end{equation}
where $T$ is given in keV. Figure~\ref{fig:dens} shows the resulting
entropy profile.
This figure shows what is becoming an ubiquitous feature of cluster
cores, an ``entropy floor'' in the central regions.
Inside of $\sim 20$ kpc, the gas is essentially isentropic.
Various mechanisms have been proposed to explain the presence 
of this entropy floor including pre-heating of the gas 
(Kaiser 1991) by supernovae-driven winds (e.g. Ponman, Cannon, \&
Navarro 1999) and active galactic nuclei (e.g. Yamada \& Fujita 2001; Nath
\& Roychowdhury 2002). Alternatively, it has been argued
that radiative cooling can be an efficient means for introducing an
entropy floor at the observed levels. Voit et al. (2002) have studied
the effect various entropy modifications on the properties of the
cluster gas.

\section{Conduction}
\label{sec:conduct}

\begin{figure}
\includegraphics[width=4.5in, angle=90]{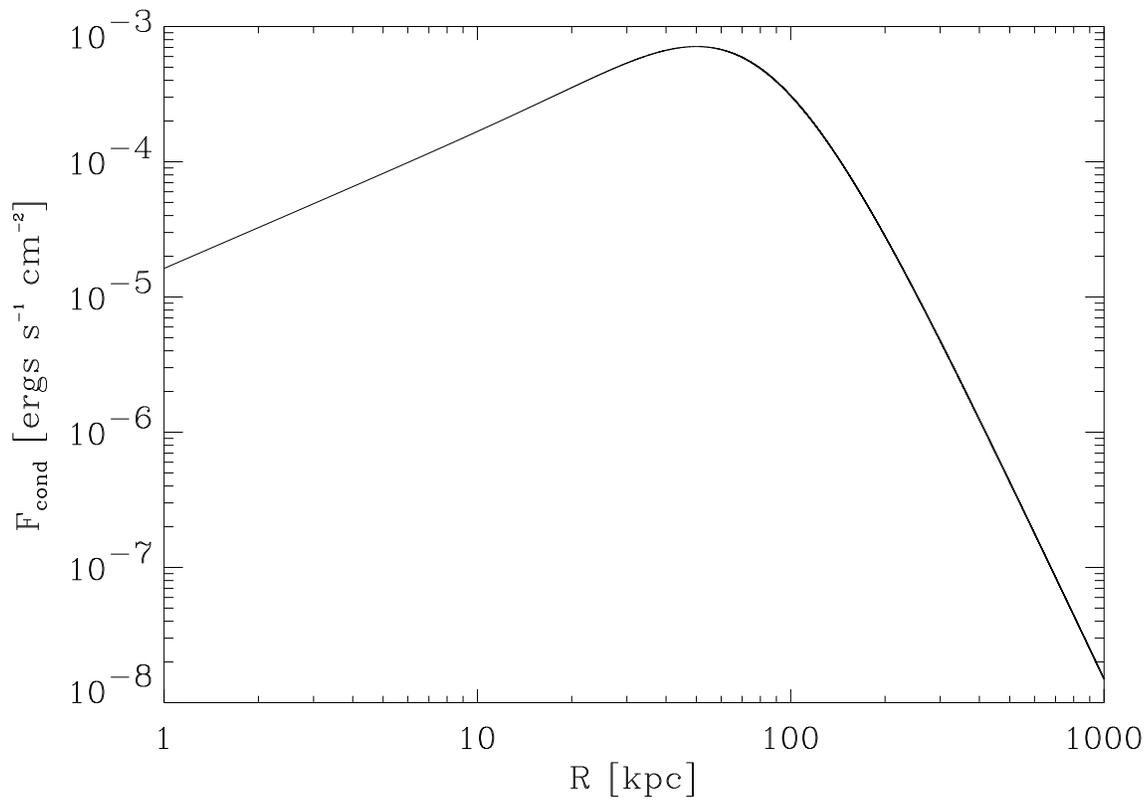}
\caption{Heat flux due to thermal conduction as a function radius
in the cluster assuming conduction operates at the Spitzer value
(i.e., $\kappa = \kappa_{sp}$).}
\label{fig:hflux}
\end{figure}

XMM RGS and \chandra HETG observations of cluster
cores  have failed to detect emission lines from gas at temperatures 
less than about 3 keV indicating that cooling may be truncated
somehow (Peterson et al. 2001, Peterson et al. 2003, Wise et al. 2004).
These data have led to the revitalization of various heating
mechanisms previously considered to be ineffective in clusters
including heating by the central AGN and thermal conduction.
Recent work by Narayan \& Medvedev (2001) and Zakamska \& Narayan (2003)
however suggests that thermal conduction may be effective in clusters.
Using the derived density and temperature in the cluster, we can
directly calculate the heat input due to thermal conduction in Abell 1068.
Following the analysis of Zakamska \& Narayan (2003), we can write
\begin{equation}
F = - \kappa \frac{dT}{dr}
\end{equation}
for the heat flux due to thermal conduction, where the thermal
conductivity is
\begin{equation}
\kappa = f \kappa_{sp} = f \left( \frac{1.84\times 10^{-5} ~T^{5/2}}
{{\rm ln} ~\Lambda} \right) 
~~{\rm ergs} ~{\rm s}^{-1} ~{\rm K}^{-1} ~{\rm cm}^{-1}
\end{equation}
and ln $\Lambda \sim 37$ is the usual Coulomb logarithm. Here, the
conductivity is expressed relative to the canonical Spitzer value,
$\kappa_{sp}$ by a fixed fraction $f$.
Figure~\ref{fig:hflux} shows the resulting conductive heat flux
as a function of radius in the cluster assuming $f=1$, i.e. assuming
thermal conduction is operating at the Spitzer value.
The heat flux peaks at a radius of $\sim 50$ kpc where the 
temperature gradient is steepest.

\begin{figure}
\includegraphics[width=4.5in, angle=90]{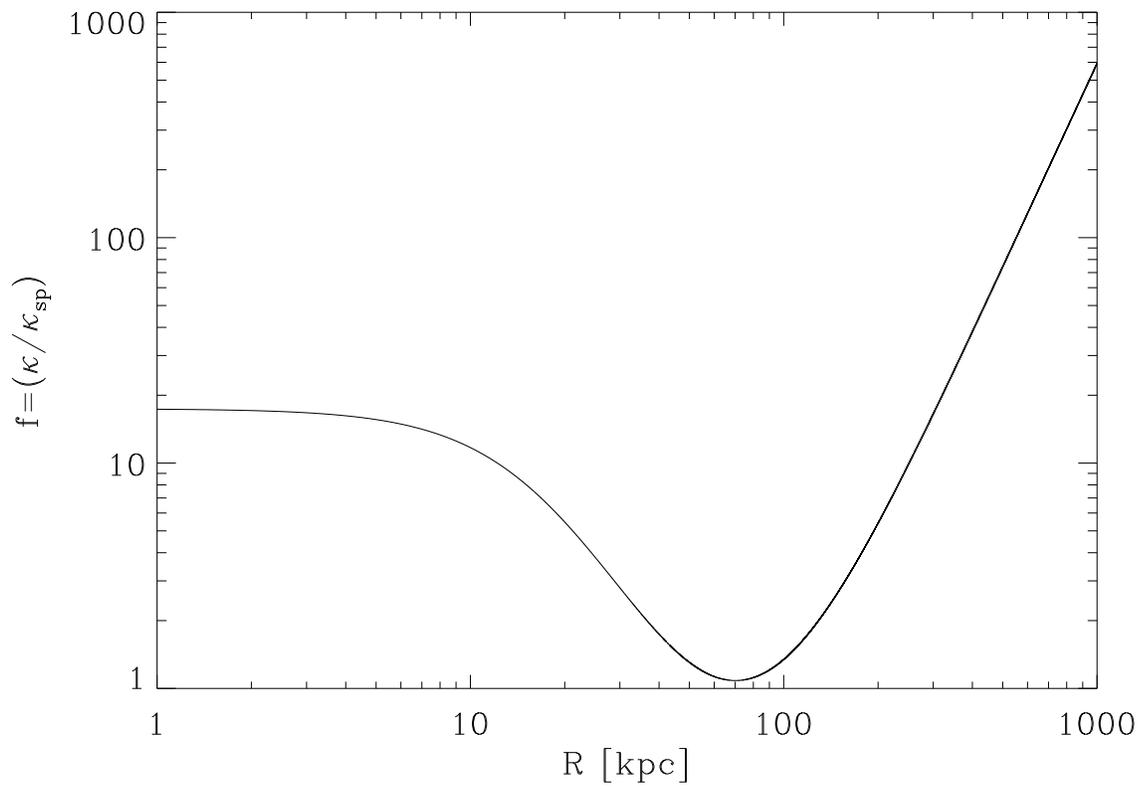}
\caption{Value of the conduction parameter, f, required to balance
radiative cooling as a function of radius in the cluster. The
conduction parameter is expressed as a ratio relative to the Spitzer value.}
\label{fig:fval}
\end{figure}

A more interesting quantity to calculate, however, is the required
level of thermal conduction necessary to balance cooling.
If we assume that the observed X--ray luminosity due to cooling
within a given radius is completely balanced by heating due to thermal
conduction from hotter gas outside that radius, we can write
\begin{equation}
L_X (<r) = 4 \pi r^2 ~f ~\kappa_{sp} \left( \frac{dT}{dr} \right).
\end{equation}
Solving this equation for the required value of the conduction
parameter, $f$, at each position in the cluster yields 
Figure~\ref{fig:fval}. As this figure shows, the thermal conductivity
of the gas must exceed the Spitzer value by an order of magnitude
to balance radiative cooling over the cooling radius in Abell 1068.
Consequently, it seems unlikely that thermal conduction could
quench cooling in Abell 1068, especially in the core.
Similar results have been found for analyses of other cooling flow
clusters observed by \chandra including Abell 2052, Abell 2597, Hydra A, Ser
159-03, and 3C 295 (Zakamska \& Narayan 2003) as well as the cores
of Abell 2199 and Abell 1835 (Voigt et al. 2002; Fabian, Voigt, \&
Morris 2002). As we show in Paper II, energy input from other sources
such as the central AGN and supernovae can account for $\leq$ 25\% of
the X--ray luminosity from cooling gas.

%
%
\section{Conclusions}
\label{sec:conclude}

We have performed a detailed spatial and spectral analysis of a 
26.8 ksec Chandra X--ray observation of the Abell 1068 cluster 
of galaxies. This cluster is exceptional due to the presence 
of a massive 20---70 M$_{\odot}$ yr$^{-1}$ starburst in the core. Our
primary goal was to determine whether or not the data for this object
are consistent with the standard cooling flow model where cooling
X--ray plasma accumulates in some cold form such as stars.
Our analysis indicates that Abell 1068 exhibits many of the 
common characteristics seen in other clusters with cool cores
including a sharply peaked surface brightness profile, declining
temperature gradient, and flat entropy profile in the core.
Although discriminating between various spectral models can
be difficult at CCD resolutions, the integrated X--ray spectrum
for Abell 1068 is best fit by a cooling flow model with
a total mass deposition rate of $\sim 150$ M$_{\odot}$ yr$^{-1}$.
Inside a radius of 40 kpc, the measured cooling rate drops
to $\sim 40$ M$_{\odot}$ yr$^{-1}$ which is completely consistent
with the star formation rates.

Although measuring mass depositions rates from X--ray data is 
notoriously difficult to interpret due to the degeneracy in spectral 
fits, the cooling time in the gas is a well constrained quantity. The
X--ray surface brightness distribution for Abell 1068 provides an
accurate measure of the density in the gas. When combined with the
measured temperature profile, we find that the cooling time of the
gas in the core of Abell 1068 is very short, reaching a minimum
of $9 \times 10^7$ yrs. The cooling time inside 40 kpc, where 98\%
of the star formation is observed in Abell 1068 occurs (Paper II),
is $\leq 5 \times 10^8$ yrs. 
The close spatial correspondence between regions of observed star
formation and short cooling times again points to a connection 
between the cooling X--ray plasma and the star forming material.

The abundance profile in Abell 1068 is especially striking.
The annular analysis clearly indicates a central enhancement
in the core roughly a factor of 2 higher than in the outer
regions. We have also constructed a 2D map of the abundance
in the cluster which shows a strong spatial correlation with
the optical light. Our analysis indicates that this central increase
is consistent with enhancement by supernova explosions associated with
the cD's ambient stellar population and the recent starburst. 
This correlation also argues against the presence of significant
energy input due to ``bubbles'' generated by a central AGN,
since such a mechanism would tend to smooth out such enhancements
on a relatively short timescale. And in any event, the central
radio source in Abell 1068 is an order of magnitude weaker than
those in clusters like Hydra A and Perseus where such bubbles
are observed.

Finally, we have examined the possibility that heating by thermal
conduction could quench cooling in this object. We find that
conduction is an order of magnitude too weak to balance the
emission from cooling over the entire cooling region. In Paper II,
we have examined various other potential feedback mechanisms
including the central AGN and energy input from supernovae.
We estimate that these mechanisms could account for $\lae 25$\% of the
cooling luminosity. 
Taken all together, these results are consistent with a picture where
the extreme starburst properties of Abell 1068 are being fueled
by a cooling ICM.

%
%
\vspace{0.25in}
\acknowledgments
The authors would like to thank John Houck for many helpful
discussions and a variety of technical support. M. Wise was 
supported by the Smithsonian Astrophysical Observatory (SAO) contract
SVI-61010 for the Chandra X-Ray Center (CXC). B. McNamara  was
supported by NASA Long Term Space Astrophysics Grant NAG5--11025
and Chandra Archival Research Grant AR2--3007X. S. Murray was
supported by NASA Grant NAS8--01130.

%
%

%
%

%
%
\begin{deluxetable}{crrrrrr}
\tablecolumns{7}
\tablewidth{0pc}
\tablecaption{Integrated Spectral Fits}
\tablehead{ 
\colhead{Model} & \colhead{$N_H$\tablenotemark{a}} & \colhead{$kT$\tablenotemark{b}} & \colhead{$A$ } & \colhead{$kT_{min}$\tablenotemark{b}} & \colhead{$\dot{\rm{M}}_X$} & \colhead{$\chi^2_{\rm norm}$}             } 
 \startdata 
      1T & $1.26_{-0.54}^{+0.45}$ & $3.76_{-0.14}^{+0.22}$ & $0.47_{-0.05}^{+0.08}$ & ~~~---~~~ & ~~~---~~~ & $1.26$ \\ 
      1T & $1.40$ & $3.76_{-0.14}^{+0.13}$ & $0.47_{-0.05}^{+0.05}$ & ~~~---~~~ & ~~~---~~~ & $1.25$ \\ 
      2T & $1.39_{-0.58}^{+0.33}$ & $6.85_{-1.92}^{+2.15}$ & $0.38_{-0.06}^{+0.07}$ & $2.16_{-0.44}^{+0.46}$ & ~~~---~~~ & $1.15$ \\ 
      2T & $1.40$ & $6.84_{-1.86}^{+2.16}$ & $0.38_{-0.06}^{+0.06}$ & $2.16_{-0.43}^{+0.45}$ & ~~~---~~~ & $1.13$ \\ 
      FC & $2.21_{-0.52}^{+0.53}$ & $4.21_{-0.20}^{+0.21}$ & $0.50_{-0.06}^{+0.06}$ & $0.10$ & $145.09_{-36.77}^{+41.87}$ & $1.12$ \\ 
      FC & $1.40$ & $4.27_{-0.22}^{+0.18}$ & $0.54_{-0.07}^{+0.05}$ & $0.10$ & $114.07_{-33.42}^{+26.05}$ & $1.14$ \\ 
      TC & $1.19_{-0.27}^{+0.59}$ & $5.92_{-1.01}^{+0.66}$ & $0.44_{-0.07}^{+0.06}$ & $2.00$ & $983.39_{-103.57}^{+131.93}$ & $1.19$ \\ 
      TC & $1.40$ & $5.96_{-0.81}^{+0.00}$ & $0.43_{-0.05}^{+0.05}$ & $2.00$ & $1038.06_{-17.55}^{+73.97}$ & $1.16$ \\ 
      VC & $1.27_{-0.41}^{+0.39}$ & $7.51_{-1.67}^{+0.03}$ & $0.40_{-0.04}^{+0.06}$ & $1.45_{-0.00}^{+0.38}$ & $790.44_{-9.20}^{+13.26}$ & $1.14$ \\ 
      VC & $1.40$ & $7.54_{-1.59}^{+0.00}$ & $0.40_{-0.04}^{+0.04}$ & $1.48_{-0.01}^{+0.35}$ & $805.63_{-0.00}^{+27.32}$ & $1.13$ \\ 
\enddata
\tablenotetext{a}{Galactic column in units of $10^{20}$ cm$^{-2}$.}
\tablenotetext{b}{Temperature in keV.}
\label{tab:spectfit}
\end{deluxetable}

%
%
\begin{deluxetable}{crrrrrrrrrr}
\rotate
\tablecolumns{11}
\tablewidth{0pc}
\tablecaption{Beta-Model Fits}
\tablehead{ 
\colhead{$\Delta$ E [keV]} & \colhead{$r_1$} [\arcsec] & \colhead{$\beta_1$}  & \colhead{$I_1$\tablenotemark{a}} & \colhead{$r_2$} [\arcsec] & \colhead{$\beta_2$}  & \colhead{$I_2$\tablenotemark{a}} & \colhead{$I_B$\tablenotemark{a}}  & \colhead{$\chi^2_{\rm norm}$}  & \colhead{$\chi^2$}  & \colhead{N}             } 
 \startdata 
0.29--7.00 & $24.19_{-1.55}^{+1.91}$ & $0.598_{- 0.006}^{+ 0.006}$ & $1294.9_{-186.08}^{+156.72}$ & $7.65_{-1.28}^{+1.53}$ & $0.990_{- 0.165}^{+ 0.198}$ & $7717.4_{-482.84}^{+527.77}$ & $2.542_{-0.04}^{+0.04}$ & 1.77 & 723.7 &  417  \\ 
0.29--2.00 & $21.27_{-1.09}^{+1.40}$ & $0.587_{- 0.005}^{+ 0.006}$ & $1239.5_{-124.34}^{+103.43}$ & $10.96_{-1.83}^{+2.19}$ & $1.665_{- 0.278}^{+ 0.333}$ & $5950.4_{-364.17}^{+404.37}$ & $1.050_{-0.03}^{+0.03}$ & 1.19 & 486.0 &  416  \\ 
2.00--7.00 & $26.37_{-3.06}^{+5.11}$ & $0.574_{- 0.015}^{+ 0.035}$ & $193.0_{-32.17}^{+38.61}$ & $4.38_{-0.73}^{+0.88}$ & $0.554_{- 0.055}^{+ 0.111}$ & $1377.6_{-219.51}^{+267.37}$ & $1.458_{-0.03}^{+0.03}$ & 1.19 & 488.6 &  417  \\ 
\enddata
\tablenotetext{a}{Surface brightness $I$ in units of $10^{-9}$ photons sec${}^{-1}$ cm${}^{-2}$ arcsec${}^{-2}$}
\label{tab:betafit}
\end{deluxetable}

%
%
\begin{deluxetable}{crrrrr}
\tablecolumns{5}
\tablewidth{0pc}
\tablecaption{Mass Estimates}
\tablehead{                
\colhead{Model}           &
\colhead{$kT$}            &
\colhead{$n_0$}           &
\colhead{$M_{gas}$}       &
\colhead{$M_{tot}$}       \\
\colhead{}                &
\colhead{[keV]}           &
\colhead{[$10^{-2}$ cm$^{-3}$]}       &
\colhead{[10$^{13} M_\odot$]}       &
\colhead{[10$^{13} M_\odot$]}       }
\startdata
      1T &    $3.76_{-0.14}^{+0.22}$  & $  2.40_{ -0.17}^{+  0.17}$ & $  4.72_{ -0.80}^{+  0.90}$ & $ 25.28_{ -1.95}^{+  1.99}$ \\
      2T &    $6.85_{-1.92}^{+2.15}$  & $  2.38_{ -0.17}^{+  0.17}$ & $  4.68_{ -0.79}^{+  0.89}$ & $ 45.45_{-11.66}^{+ 12.08}$ \\
      FC &    $4.21_{-0.20}^{+0.21}$  & $  2.44_{ -0.18}^{+  0.17}$ & $  4.81_{ -0.82}^{+  0.91}$ & $ 27.41_{ -2.24}^{+  2.31}$ \\
      TC &    $5.92_{-1.01}^{+0.66}$  & $  2.37_{ -0.17}^{+  0.17}$ & $  4.67_{ -0.79}^{+  0.89}$ & $ 41.21_{ -7.27}^{+  7.63}$ \\
    Core &    $1.76_{-0.10}^{+0.21}$  & $ 13.53_{ -1.38}^{+  1.56}$ & $  0.15_{ -0.10}^{+  0.32}$ & $ 19.77_{ -4.68}^{+  5.64}$ \\
\enddata
\label{tab:mass}
\end{deluxetable}

%
%
\clearpage

%
%
\clearpage

%
%
\clearpage

%
%
\clearpage

%
%
\clearpage

%
%
\clearpage

%
%
\clearpage

%
%
\clearpage

%
%
\clearpage

%
%
\clearpage

%
%
\clearpage 

%
%
\clearpage 

%
%
\clearpage 

%
%
\clearpage 

%
%
\clearpage 

%
%

\begin{thebibliography}{}

\bibitem[Allen, Ettori, \& Fabian 2001]{allen01} Allen, S. W., Ettori, S.,
         Fabian, A. C. 2001, MNRAS, 324, 877.

\bibitem[Arnaud 1996]{xspec} Arnaud, K.A., 1996, Astronomical Data
         Analysis Software and Systems V, eds. Jacoby G. and 
         Barnes J., p17, ASP Conf. Series volume 101. 

\bibitem[Brighenti \& Mathews 2002]{brighenti02} Brighenti, F. \&
         Mathews, W. G. 2002, ApJ, 574, L11.

\bibitem[Cardiel et al. 1998]{cardiel98} Cardiel, N., Gorgas, J., \& 
         Aragon-Salamanca, A. 1998, MNRAS, 298, 977.

\bibitem[Crawford et al. 1999]{crawford99} Crawford, C. S.,
         Allen, S. W., Ebeling, H., Edge, A. C., Fabian, A. C.
         1999, MNRAS, 306, 857.

\bibitem[David et al. 2001]{david01}David, L. P., Nulsen, P. E. J.,
         McNamara, B. R., Forman, W., Jones,  C., Ponman, T., 
         Robertson, B., Wise, M. 2001, ApJ, 557, 546.

\bibitem[Donahue et al. 2000]{donahue00} Donahue, M., Mack, J., 
         Voit, G. M., Sparks, W. B., Elston, R., \& Maloney, P. R. 2000, 
         ApJ, 545, 670.

\bibitem[Edge 2001]{edge01} Edge, A. C. 2001, MNRAS, 328,762.

\bibitem[Ettori 2000]{dbeta2} Ettori, S. 2000 MNRAS, 318, 1041.

\bibitem[Fabian 1994]{fabian94} Fabian, A. C. 1994, ARAA, 32, 277.

\bibitem[Fabian et al. 2000]{fabian00} Fabian, A. C., Mushotzky,
         R. F., Nulsen, P. E. J., Peterson, J. R. 2000, MNRAS, 321, L20. 

\bibitem[Fabian, Voigt, \& Morris 2002]{fabian02} Fabian, A. C.,
         Voigt, L. M., Morris, R. G. 2002, MNRAS, 335, 71. 

\bibitem[Falcke et al. 1998]{falcke98} Falcke, H., Rieke, M. J., 
         Rieke, G. H., Simpson, C., Wilson, A. 1998, ApJ, 494, 155.

\bibitem[Fukazawa et al. 2000]{fukazawa00} Fukazawa, Y., Makishima,
         K., Tamura, T., Nakazawa, K., Ezawa, H., 
         Ikebe, Y., Kikuchi, K., \& Ohashi, T. 2000, MNRAS, 313, 21.    

\bibitem[Hicks et al. 2002]{hicks2002} Hicks, A. K., Wise, M. W.,
  Houck, J. C., Canizares, C. R. 2002, ApJ, 580, 763.

\bibitem[Houck \& DeNicola 2000]{isis} Houck,~J.~C.~\& Denicola,
         L.~A.\ 2000, ASP Conf.~Ser.~216: 
         Astronomical Data Analysis Software and Systems IX, 9, 591.

\bibitem[Houck et al. 2004]{tmap} Houck, J.C., Davis, D. S., Wise,
         M. W. 2004, ApJ, in preparation. 

\bibitem[Irwin \& Bregman 2001]{irwin01} Irwin, J. A., \& Bregman,
         J. N.,  2001, ApJ, 546, 150.

\bibitem[Jaffe et al. 1997]{jaffe97} Jaffe, W. \& Bremer, M. N. 1997, 
         MNRAS, 284, L1.

\bibitem[Jaffe et al. 2001]{jaffe01} Jaffe, W. \& Bremer, M. N., 
         \& van der Werf, P. P. 2001, MNRAS, 324, 443.

\bibitem[Johnstone et al. 2002]{johnstone02} Johnstone, R. M., Allen, S. W.,
         Fabian, A. C., Sanders, J. S. 2002, in press (astro-ph/0202071).

\bibitem[Kaiser 1991]{kaiser91} Kaiser, N. 1991, ApJ, 383, 104.

\bibitem[Markevitch et al. 2000]{markevitch00} Markevitch, M., Ponman,
         T. J., Nulsen, P. E. J., Bautz, M. W., Burke, D. J., David,
         L. P., Davis, D., Donnelly, R. H., Forman, W. R., 
         Jones, C., Kaastra, J., Kellogg, E., Kim, D.-W., Kolodziejczak, J.,
         Mazzotta, P., Pagliaro, A., Patel, S., Van Speybroeck, L., 
         Vikhlinin, A., Vrtilek, J., Wise, M., \& Zhao, P.
         2000, ApJ, 541, 542.

%


\bibitem[McNamara 2002]{mcnamara02} McNamara, B. R. 2002, 
             in ``The High Energy Universe at Sharp The 
             High Energy Universe at Sharp Focus: Chandra Science,'' 
             San Francisco: ASP Conf. Ser., 262, 351, eds. S. Vrtilek,
             E. M. Schlegel, \& L. Kuhi, astro-ph/0202199.

\bibitem[McNamara \& O'Connell 1989]{mcnamara89} McNamara, B. R. 
         \& O'Connell, R. W. 1989, AJ, 98, 2018.

\bibitem[McNamara et al. 2000]{mcnamara00} McNamara, B. R., Wise, M.,
         Nulsen, P. E. J., David, L. P., Sarazin,  
         C. L., Bautz, M., Markevitch, M., Vikhlinin, A., Forman,
         W. R., Jones, C., \&  Harris, D. E. 2000, ApJ, 534, L135.

\bibitem[McNamara, Wise, \& Murray 2004]{paper2} McNamara, B. R., 
         Wise, M. W., Murray, S. S. 2004, ApJ, in press.

\bibitem[Nath \& Roychowdhury 2002]{nath02} Nath, B. B. 
         \& Roychowdhury, S. 2002, MNRAS, 333, 145.

\bibitem[Narayan \& Medvedev 2001]{narayan01}Narayan, R. \& Medvedev, M.V. 
         2001, ApJ, 562, 129.

\bibitem[Nulsen et al. 2002]{nulsen02} Nulsen, P. E. J., David, L. P.,
         McNamara, B. R., Jones, C., Forman,  
         W.R., \& Wise, M. 2002, ApJ, 568, 163.

\bibitem[Peletier et al. 1990]{peletier90} Peletier, R. F., Davies,
         R. L., Illingworth, G. D., Davis, L. E., \& Cawson, M. 
         1990, AJ, 100, 1091.

\bibitem[Peterson et al. 2001]{peterson01} Peterson, J. R., Paerels, F. B. S., 
         Kaastra, J. S., Arnaud, M., Reiprich, T. H., Fabian, A. C.,
         Mushotzky, R. F., Jernigan, J. G., Sakelliou, I. 2001, A\&A, 365, 104.

\bibitem[Peterson et al. 2003]{peterson03} Peterson, J. R., Kahn,
         S. M., Paerels, F. B. S., Kaastra, J. S., 
         T. Tamura, J. A. M., Bleeker, \& C. Ferrigno 2003, ApJ,
         submitted, astro-ph/0210662.

\bibitem[Ponman, Cannon, \& Navarro 1999]{ponman99} Ponman, T. J.,
         Cannon, D. B., \& Navarro, J. F. 1999, Nature, 397, 135.

\bibitem[Ruszkowski \& Begelman 2002]{ruszkowski02} Ruszkowski, M. 
         \& Begelman, M. C. 2002, ApJ, 581, 223.

\bibitem[Smith et al. 2001]{apec} Smith, R. K., Brickhouse, N. S.,
         Liedahl, D. A., Raymond, J. C. 2001, ApJ, 556, L91.

\bibitem[Tammann 1974]{tammann74} Tammann, G. A. 1974, in 
         Supernovae and Supernova Remnants, Proceedings of the
         International Conference on Supernovae, 
         Dordrecht: Reidel, ed. by 
         C. B. Cosmovici, Astrophysics and Space Science Library, 
         Vol. 45, 155.

\bibitem[Vikhlinin, Markevitch, \& Murray 2001]{vikhlinin01}
         Vikhlinin, M., Markevitch, M, Murray, S. S. 2001, ApJ, 551, 160.

\bibitem[Voigt et al. 2002]{voigt02} Voigt, L. M., Schmidt, R. W.,
         Fabian, A. C., Allen, S. W., Johnstone, R. M., MNRAS, 335, 7. 

\bibitem[Voit et al. 2002]{voit02} Voit, G. M., Bryan, G. L.,
          Balogh, M. L., \& Bower, R. G. 2002, ApJ, 576, 601.

\bibitem[Wise et al. 2004]{wise_hetg} Wise, M. W., Houck, J. C.,
         Canizares, C. R., Hicks, A. K.  2004, in preparation.

\bibitem[Wise \& Houck 2004]{deproj} Wise, M. W., Houck, J. C. 2004,
         in preparation.

\bibitem[Xue \& Wu 2000]{dbeta} Xue, Y.-J., Wu, X.-P., 2000 MNRAS,
         318, 715.

\bibitem[Yamada \& Fujita 2001]{yamada01} Yamada, M. \& Fujita, Y. 
         2001, ApJ, 553, 145.

\bibitem[Zakamska \& Narayan 2003]{zakamska03} Zakamska, N. L., 
         \& Narayan, R. 2003, ApJ, 582, 162.

\end{thebibliography}
\end{document}